\documentclass[artical]{aa}

\usepackage{natbib}
\usepackage{txfonts}
\usepackage{rotating}
\usepackage{subfigure}
\usepackage{booktabs}
\usepackage[breaklinks,colorlinks,citecolor=blue,linkcolor=blue]{hyperref}
\usepackage{lscape}
\usepackage{CJK}
\graphicspath{{./figure/}}

\begin{document}

\begin{CJK}{UTF8}{gbsn}
    \title{Carbon-chain molecule survey toward four low-mass molecular outflow sources}
    \author{C. Zhang (张超)
    \inst{1}
    \inst{2}
    \and
    Y. Wu
    \inst{2}
    \inst{3}
    \and
    X.-C. Liu (刘训川)
    \inst{2}
    \inst{3}
    \and
    Mengyao Tang
    \inst{1}
    \and
    Di Li
    \inst{4}
    \inst{5}
    \and
    Jarken Esimbek
    \inst{6}
    \inst{7}
    \and
    S.-L. Qin
    \inst{1}
    }

\institute{Department of Astronomy, Yunnan University, and Key Laboratory of Astroparticle Physics of Yunnan Province, Kunming 650091, China
   \and
   Department of Astronomy, School of Physics, Peking University, Beijing, 100871, China
       \email{ywu@pku.edu.cn}
   \and
   Kavli Institute for Astronomy and Astrophysics, Peking University, Beijing 100871, China
   \and
   CAS Key Laboratory of FAST, National Astronomical Observatories, CAS, Beijing 100012, China
   \and
   NAOC-UKZN Computational Astrophysics Centre (NUCAC), University of KwaZulu-Natal, Durban 4000, South Africa
   \and
   Xinjiang Astronomical Observatory, Chinese Academy of Sciences, Urumqi 830011, China
   \and
   Key Laboratory of Radio Astronomy, Chinese Academy of Sciences, Urumqi 830011, China
       }

\abstract{}
\abstract{
We performed a carbon-chain molecule (CCM) survey toward four low-mass outflow sources, IRAS 04181+2655 (I04181), HH211, L1524, and L1598, using the 13.7 m telescope at the Purple Mountain Observatory (PMO) and the 65 m Tian Ma Radio telescope at the Shanghai Observatory. We observed the following hydrocarbons (C$_2$H, C$_4$H, c--C$_3$H$_2$), HC$_{\rm 2n+1}$N (n=1,2), C$_{\rm n}$S (n=2,3), and SO, HNC, N$_2$H$^+$.  Hydrocarbons and HC$_3$N were detected in all the sources, except for L1598, which had a marginal detection of C$_4$H and a non-detection of HC$_3$N (J=2--1). HC$_5$N and CCCS were only detected in I04181 and L1524, whereas SO was only detected in HH211. L1598 exhibits the lowest detection rate of CCMs and is generally regarded to be lacking in CCMs source. The ratio of N(HC$_3$N/N(N$_2$H$^+$)) increases with evolution in low-mass star-forming cores. I04181 and L1524 are carbon-chain-rich star-forming cores that may possibly be characterized by warm carbon-chain chemistry. In I04181 and L1524, the abundant CCCS can be explained by shocked carbon-chain chemistry. In HH211, the abundant SO suggests that SO is formed by sublimated S$^+$. In this study, we also mapped HNC, C$_4$H, c--C$_3$H$_2$, and HC$_3$N with data from the PMO. We also find that HNC and NH$_3$ is concentrated in L1524S and L1524N, respectively. Furthermore, we discuss the chemical differences between I04181SE and I04181W. The co-evolution between linear hydrocarbon and cyanopolyynes can be seen in I04181SE.
}

\titlerunning{CCMs in four outflows}
\authorrunning{Zhang et al.}

\maketitle

\section{Introduction}
Stars form in molecular cores that are cold enough to have temperatures below 10 K and dense enough to be up to 10$^{5}$ cm$^{-3}$ in volume density. These cores, known as prestellar cores \citep{1994MNRAS.268..276W,2016MNRAS.463.1008W}, have low Jeans mass and they are prone to further collapse into stars. The surrounding gas and dust grain are heated up by the embedded protostar and can be affected by the feedback of star formation processes. Thus, the molecular chemistry between prestellar cores and star-forming cores is expected to vary. Previous studies of carbon-chain molecules (CCMs) have mainly been focused on the prestellar core phase \citep{1992ApJ...392..551S,2004ApJ...617..399H,2006ApJ...646..258H,2009ApJ...699..585H}. The chemistry of CCMs in low-mass star-forming cores needs to be explored.

It was previously thought that CCMs are abundant in the early stage dark molecular cores and absent in the star-forming cores \citep{1992ApJ...392..551S,2009ApJ...699..585H}. The absence of CCMs are caused by the gas phase destruction and the depletion onto dust grains.
A new type of chemistry, that is, warm carbon-chain chemistry (WCCC), was proposed based on the detection of  high-excitation lines of CCMs in the lukewarm corino of low-mass protostar L1527, and Lupus I-1 (IRAS 15398-3359) \citep{2008ApJ...672..371S,2009ApJ...697..769S}. In the gas phase, CH$_4$ can evaporate and then react with C$^+$ to efficiently form CCMs in a warm dense region heated by the emerging protostar \citep{2008ApJ...681.1385H}.
In addition, abnormally high abundances of C$_3$S were recently found in L1251A, L1221, and IRAS 20582+7724 \citep{2019MNRAS.488..495W}. Shocks can increase the abundance of S$^+$ in the gas phase and thus drive the generation of S-bearing CCMs, which is referred to as shocked carbon-chain chemistry \citep[SCCC,][]{2019MNRAS.488..495W}. Finding other low-mass star-forming cores with abundant CCMs is key to improving our understanding of the CCM chemistry in these cores.

In this paper, we report a survey of CCMs of hydrocarbons (C$_2$H, C$_4$H, c--C$_3$H$_2$), HC$_{\rm 2n+1}$N (n=1,2), C$_{\rm n}$S (n=2,3), and three dense gas tracers of SO, HNC, and N$_2$H$^+$ toward four sources, IRAS 04181+2655 (I04181), HH211, L1524, and L1598, drawn from a catalog of high-velocity molecular outflows \citep[and references therein]{2004A&A...426..503W}. The aim was to probe the chemistry evolution of CCMs in the gas phase of these regions.
The IRAS I04181 source is a Class I low-mass protostar and the associated molecule core shows bipolar outflow in CO (J=2--1) \citep{1996A&A...311..858B}. The jet-driven molecular outflow of HH211 was reported by \citet{1999A&A...343..571G}. The bipolar outflow in Class 0 protostar HH211 detected in CO (J=2--1) is compact and does not extend beyond the H$_2$ jet \citep{1994ApJ...436L.189M,2011ApJ...736...25F}.
L1524 contains Haro 6-10, which has an infrared companion \citep[Class I protostar IRAS 04263+2426;][]{1953ApJ...117...73H,1998MNRAS.299..789C}.
L1598 contains a bipolar HH complex (HH 117 and HH 118) centered on the IRAS 05496+0812 source \citep[Class I protostar,][]{2015A&A...584A..92M}. The dynamical age of L1598 is $\sim 2.2 \times10^4$ yr \citep{1988ApJ...327..350S,2002ApJS..141..157Y}. The outflow motions, as in the case of the optical jet, can heat the surrounding gas and also can produce shocks. Therefore, the four sources are good targets for probing the chemistry of CCMs. Table \ref{samples} summarizes the properties of our four target sources.

This paper is organized as follows. The observations are introduced in Sect. \ref{obs}. The results of the molecular line observations are presented in Sect. \ref{results}. We discuss the column density ratio of CCMs, N-bearing inorganic molecules, and the chemistry of CCMs in Sect. \ref{discuss}. We summarize our result in Sect. \ref{sum}.

\section{Observations \label{obs}}
\subsection{Observations with the PMO 13.7 m telescope}
Observations were carried out with the 13.7-m telescope of the Purple Mountain Observatory (PMO) from May 25 to May 29 and from Jun 1 to Jun 6, 2017. The $3\times3$-beam sideband separation Superconduction Spectroscopic Array Receiver system was used as the frontend \citep{2012ITTST...2..593S}. The HPBW is $52''$ in the 90 GHz band (3 mm). The mean beam efficiency ($\eta$) is about $50\%$. The pointing and tracking accuracies are both better than $5''$. The fast Fourier transform spectrometers (FFTSs) were used as the backend. Each FFTS with a total bandwidth of 1 GHz provides 16384 channels. The spectral resolution is about 61 kHz. The velocity resolution is $\sim$0.21 km s$^{-1}$. The upper sideband (USB) was set in 90--91 GHz with the system temperature (T$_{sys}$) around 210 K and the typical rms 0.06 K for the antenna temperature (T$_a$). The lower sideband (LSB) was set in 85--86 GHz with T$_{sys}$ around 120 K and the typical rms 0.08 K for T$_a$. Another observational frequency set is 92--93 GHz in USB and 87--88 GHz in LSB. The T$_{sys}$ and the typical rms are similar to the first setup.

The mapping observations were carried out with the on-the-fly (OTF) mode with LSB of 85-86 GHz and USB of 90-91 GHz from Aug. 31 to Sept. 7, 2017. The antenna continuously scanned a region of $18'\times 18'$ centered on the sources with a scanning rate of $30''$ per second. Because of the high rms noise level at the edges of OTF maps, only data within the center $10'\times 10'$ regions were used for analyses in this work.

\subsection{Observations with the SHAO 65 m telescope}
The spectral lines at Ku band were observed with the Tian Ma Radio Telescope (TMRT) of Shanghai Observatory on Jul. 25, 2016. The TMRT is a fully steerable radio telescope located in the west suburb of Shanghai \citep{2016ApJ...824..136L}. The frontend is a cryogenically cooled receiver covering the frequency range of 11.55-18.5 GHz. The pointing accuracy is better than 10$''$. An FPGA-based spectrometer based upon the design of Versatile GBT Astronomical Spectrometer (VEGAS) was employed as the Digital backend system (DIBAS) \citep{2012AAS...21944610B}. For molecular line observations, DIBAS supports a variety of observing modes, including 19 single sub-band modes and 10 eight sub-band modes. The center frequency of the subband is tunable to an accuracy of 10 kHz. In our observations, the DIBAS mode 22 was adopted. Each of the side-band has eight sub-bands within 1 GHz, while each sub-band has a bandwidth of 23.4 MHz and 16384 channels. The observations of L1598 were carried out with two side-bands. The observations of the other three sources were carried out with three side-bands. As shown in Table \ref{transitions}, only SO (J=1$_2$--1$_1$) was not observed in L1598. The main beam efficiency ($\eta$) is 60\% at the Ku band \citep{2015AcASn..56...63W,2016ApJ...824..136L}.
The observed lines are listed in Table \ref{transitions}. The data were reduced by CLASS and GREG in the IRAM software package GILDAS \citep{2000ASPC..217..299G,2005sf2a.conf..721P} and analysed by the open-source Python package.

\section{Results \label{results}}

\subsection{Spectral lines}
\begin{figure*}
  \centering
    \includegraphics[width=0.32\linewidth]{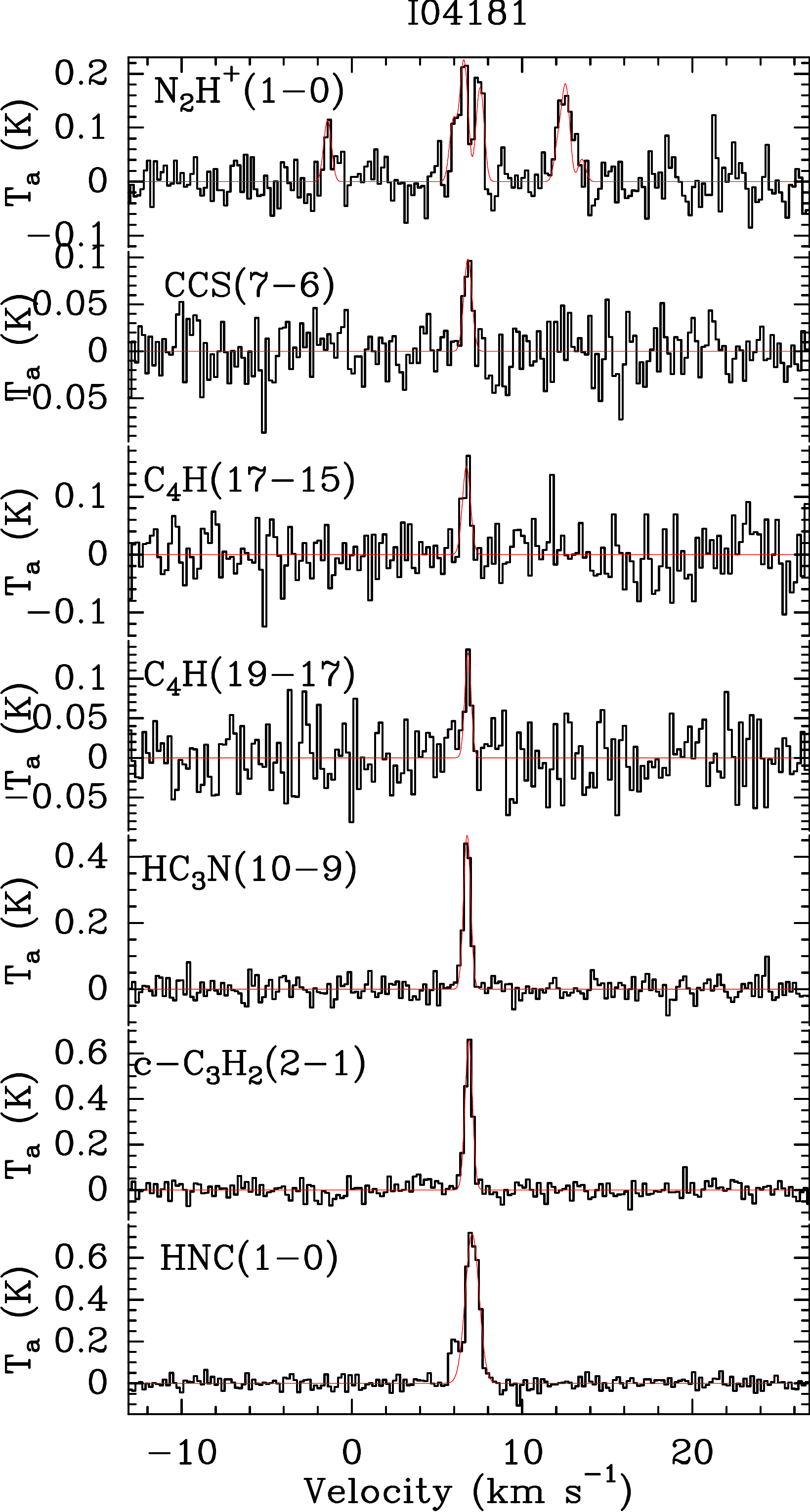}
    \includegraphics[width=0.32\linewidth]{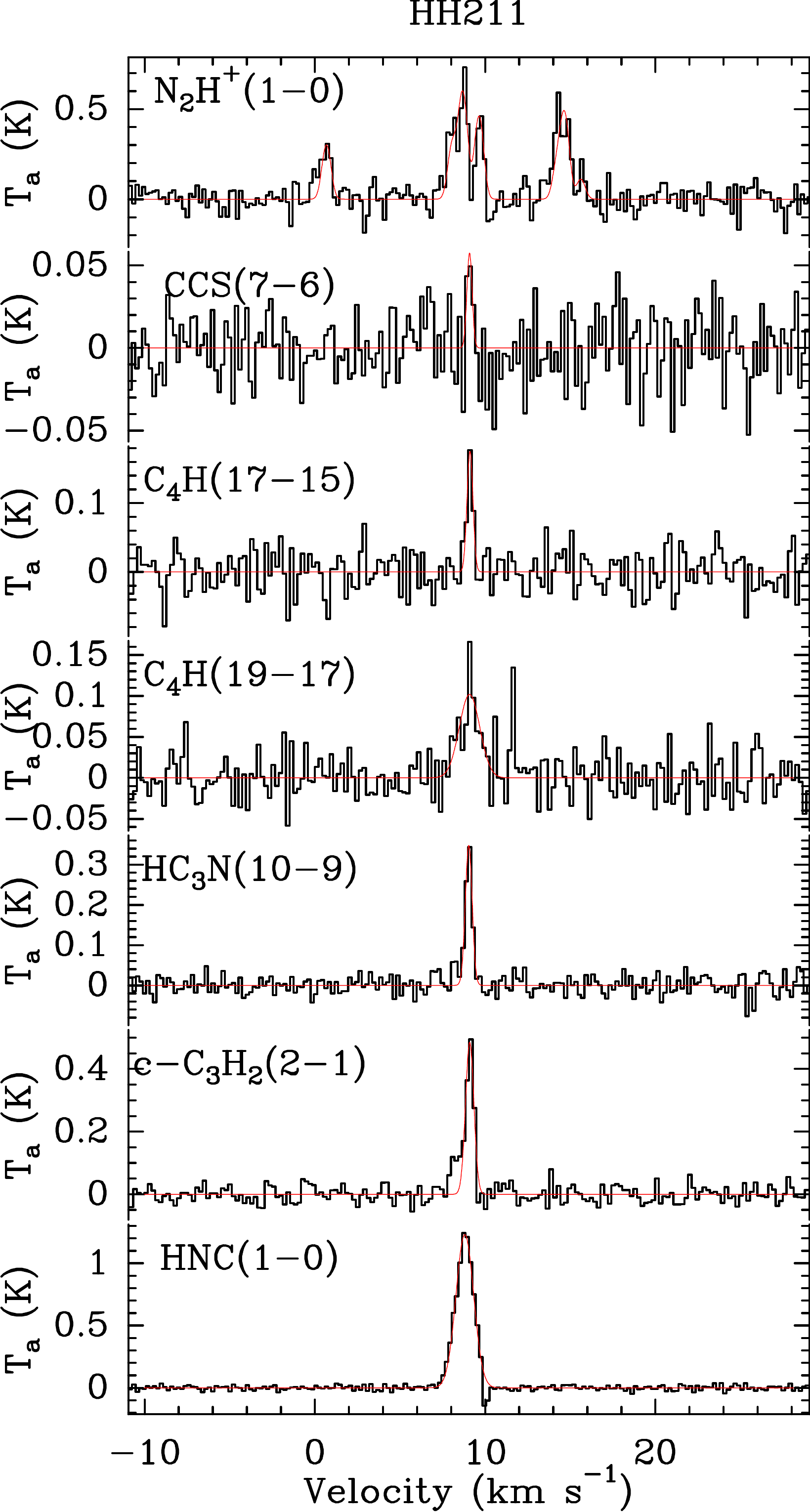}\\
    \includegraphics[width=0.32\linewidth]{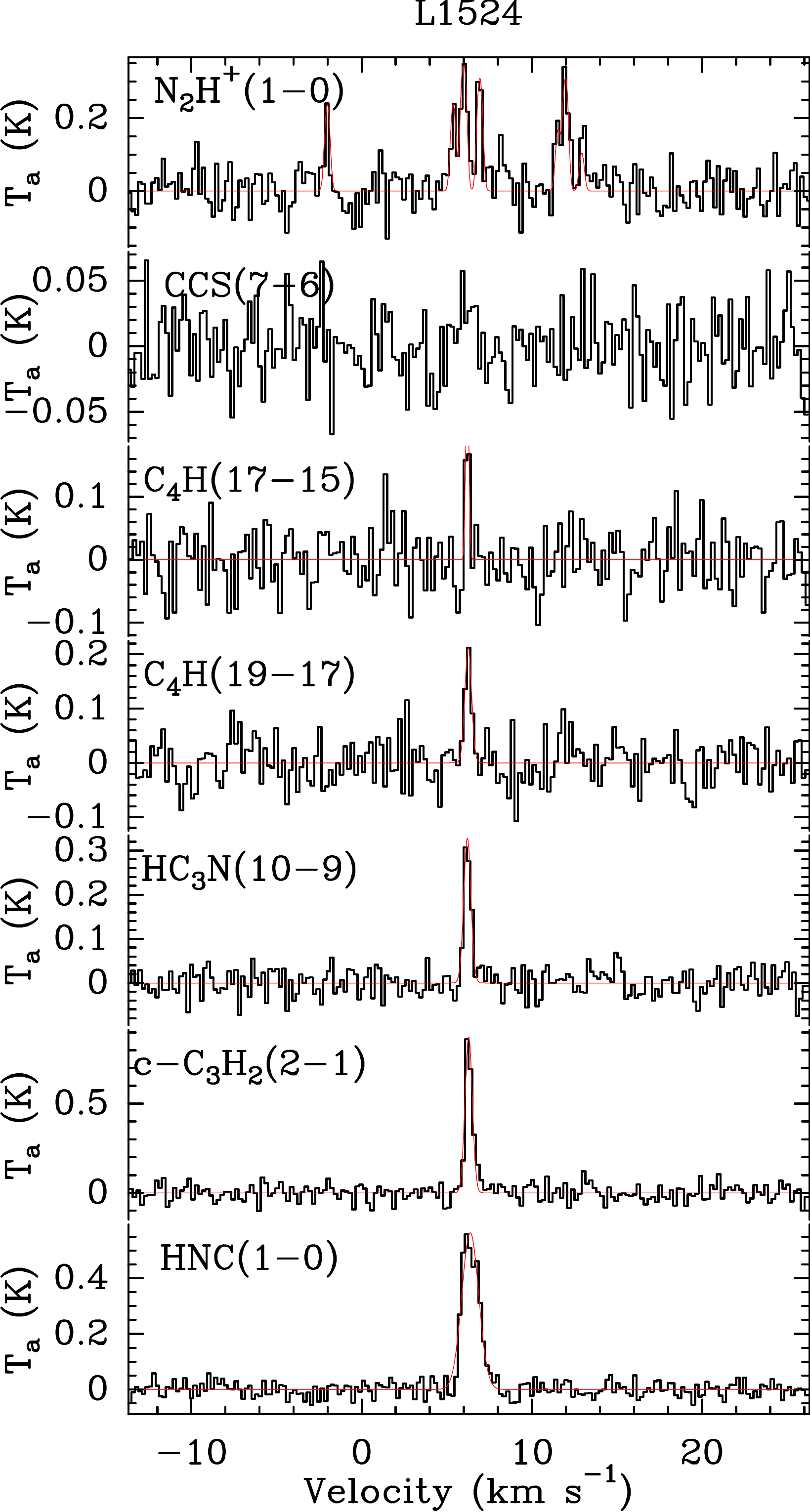}
    \includegraphics[width=0.32\linewidth]{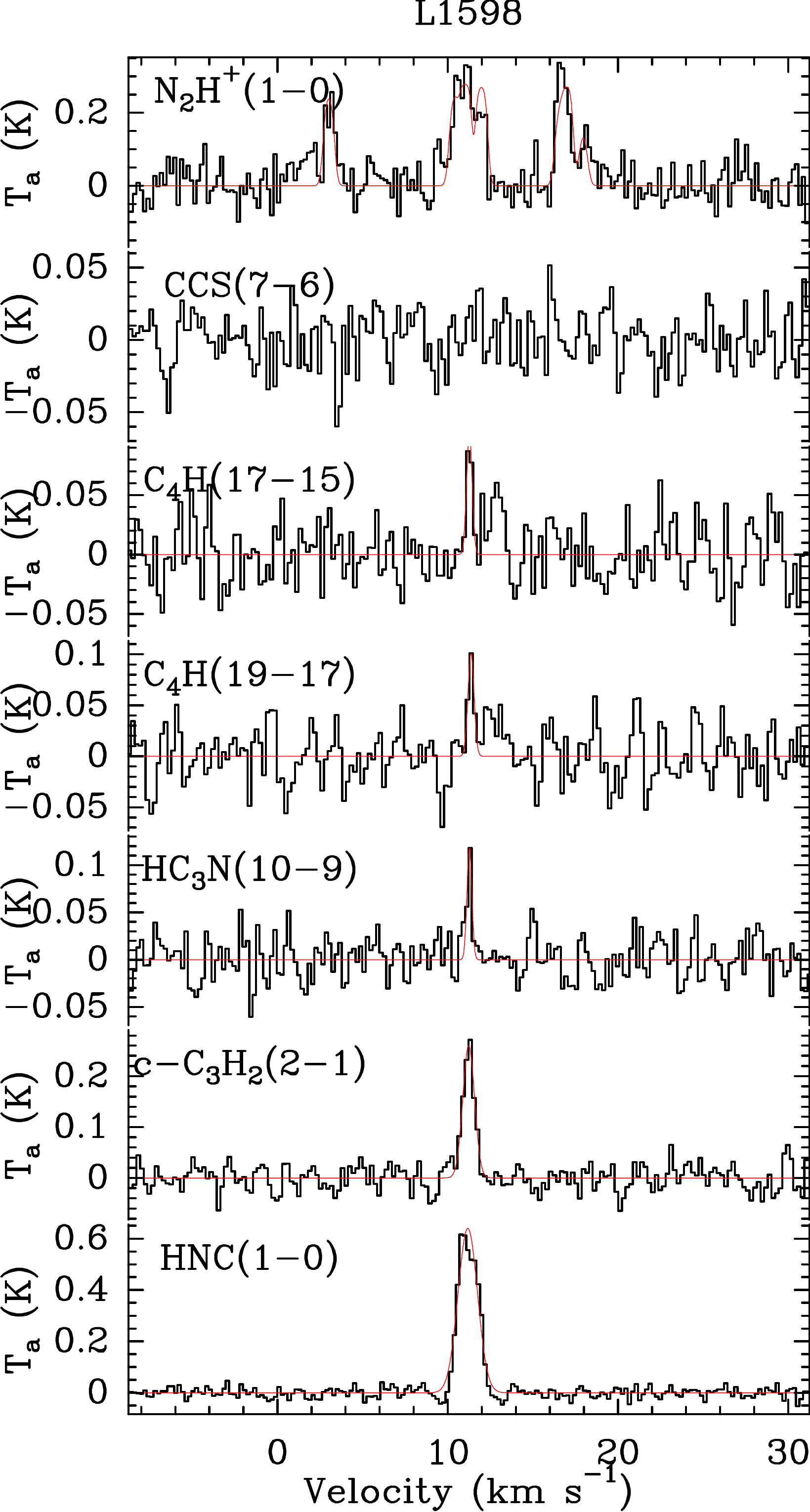}
  \caption{Spectra observed by PMO. The fitting results in the color red are overlayed on each spectrum. The molecular lines are labeled at the upper left of each spectrum. The source name is labeled at the top of each panel. \label{PMO}}
\end{figure*}

\begin{figure*}[htp!]
  \centering
  \includegraphics[width=0.8\linewidth]{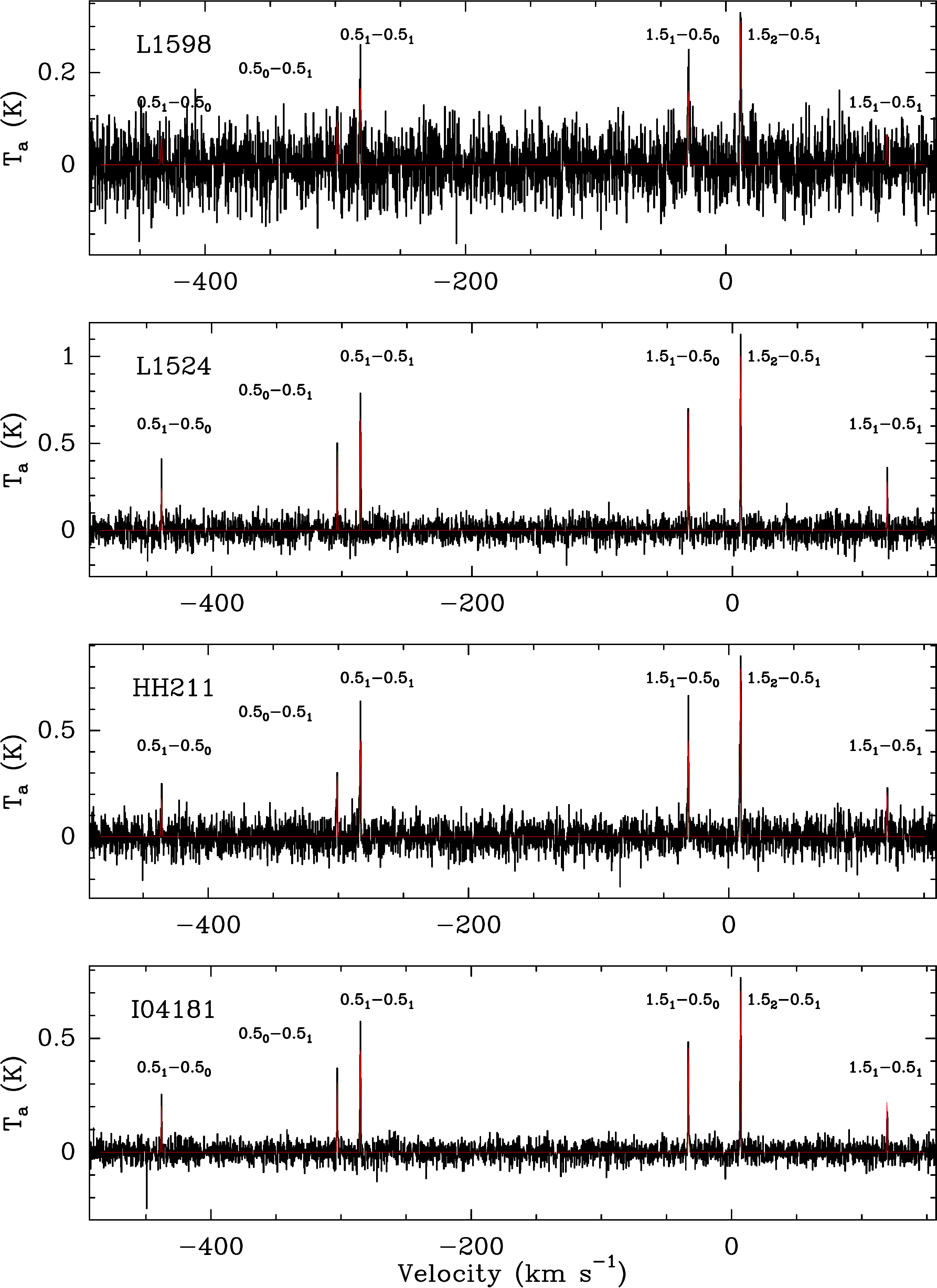}
  \caption{Spectra of CCH (N=1--0) observated by PMO. The HFS fitting results in color red are overlayed on each spectrum. The source name is labeled at upper left of spectrum. The hyperfine structures are marked. \label{appendix-cch}}
\end{figure*}

In the 3 mm band, CCH (N=1--0), N$_2$H$^+$ (J=1--0), c--C$_3$H$_2$ (J=2$_{1,2}$--1$_{0,1}$), HNC (J=1--0), and HC$_3$N (J=10--9) are detected in all the target sources, except for L1598, with non-detections of the transitions J=3/2--1/2 F=1--1 and J=1/2--1/2 F=0--1, F=1--0 of CCH (N=1--0). CCS (J=7--6) is detected with signal-to-noise ratio (S/N) of 5 in I04181 and is tentatively detected with S/N of 3 in HH211. The S/N is defined as T$_a$/T$_{rms}$. The transitions J=17/2--15/2 and J=19/2--17/2 of C$_4$H are detected with S/N larger than 3 in I04181, HH211, and L1524 and are tentatively detected with S/N $\sim$2.5 in L1598. The spectral lines of I04181, HH211, L1524, and L1598 are plotted in Fig. \ref{PMO}. 
All six hyperfine structure (HFS) components are well resolved for CCH (N=1--0) of the detected sources except for L1598. The CCH spectra are shown in Fig. \ref{appendix-cch}. The HFS components of N$_2$H$^+$ (J=1--0) are classified as three groups labeled as F$_1$=1--1, F$_1$=2--1, and F$_1$=0--1. The group F$_1$=1--1 and the group F$_1$=2--1 each contains three blended HFS.

\begin{figure*}
  \centering
  \includegraphics[width=0.8\linewidth]{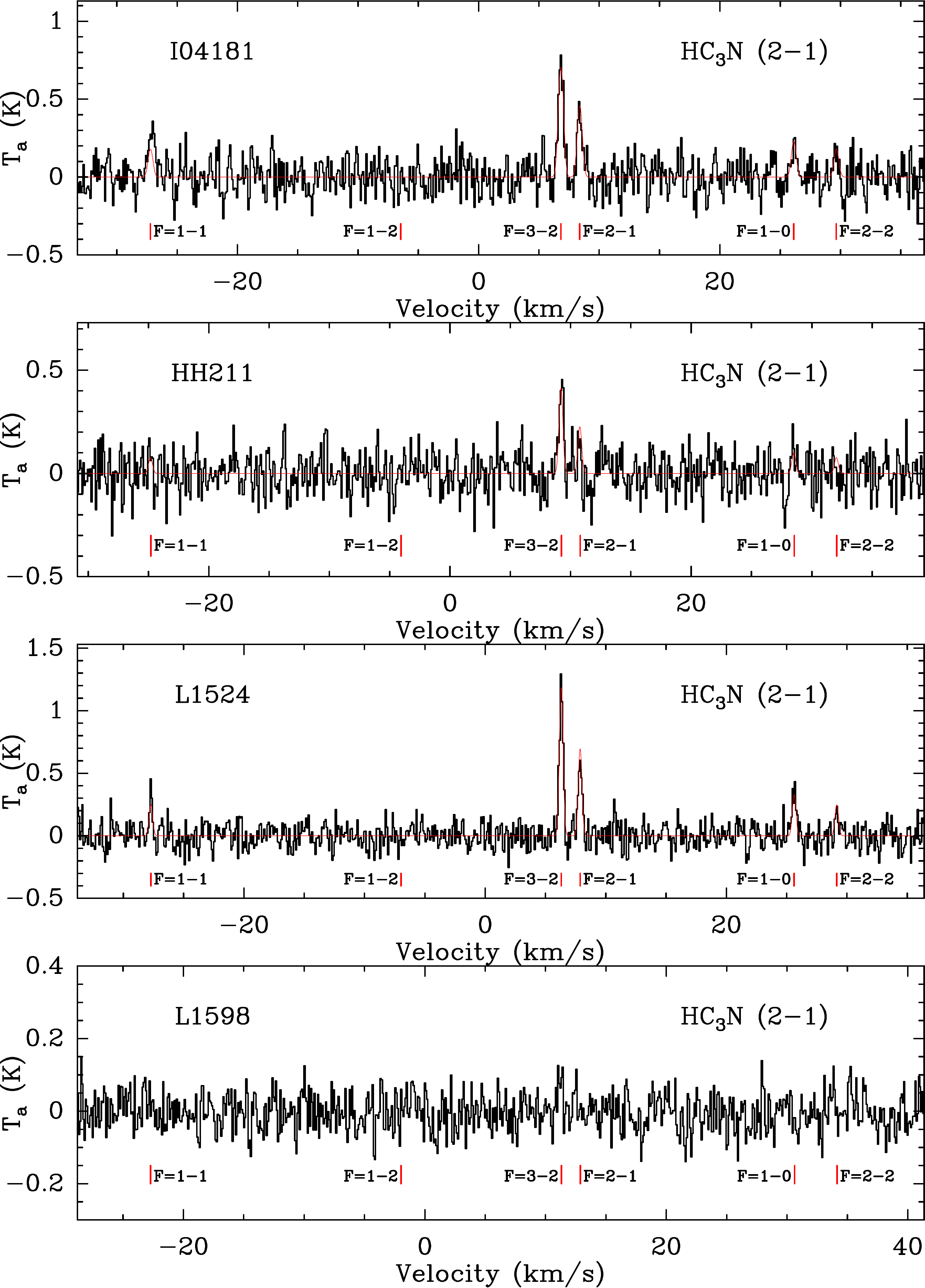}
  \caption{Spectra of HC$_3$N in transition J=2-1 observed by TMRT. The source name is labeled at the upper left of each spectrum. The hyperfine structures are marked below the spectrum. \label{HC3N}}
\end{figure*}

\begin{figure*}
  \centering
  \includegraphics[width=0.24\linewidth]{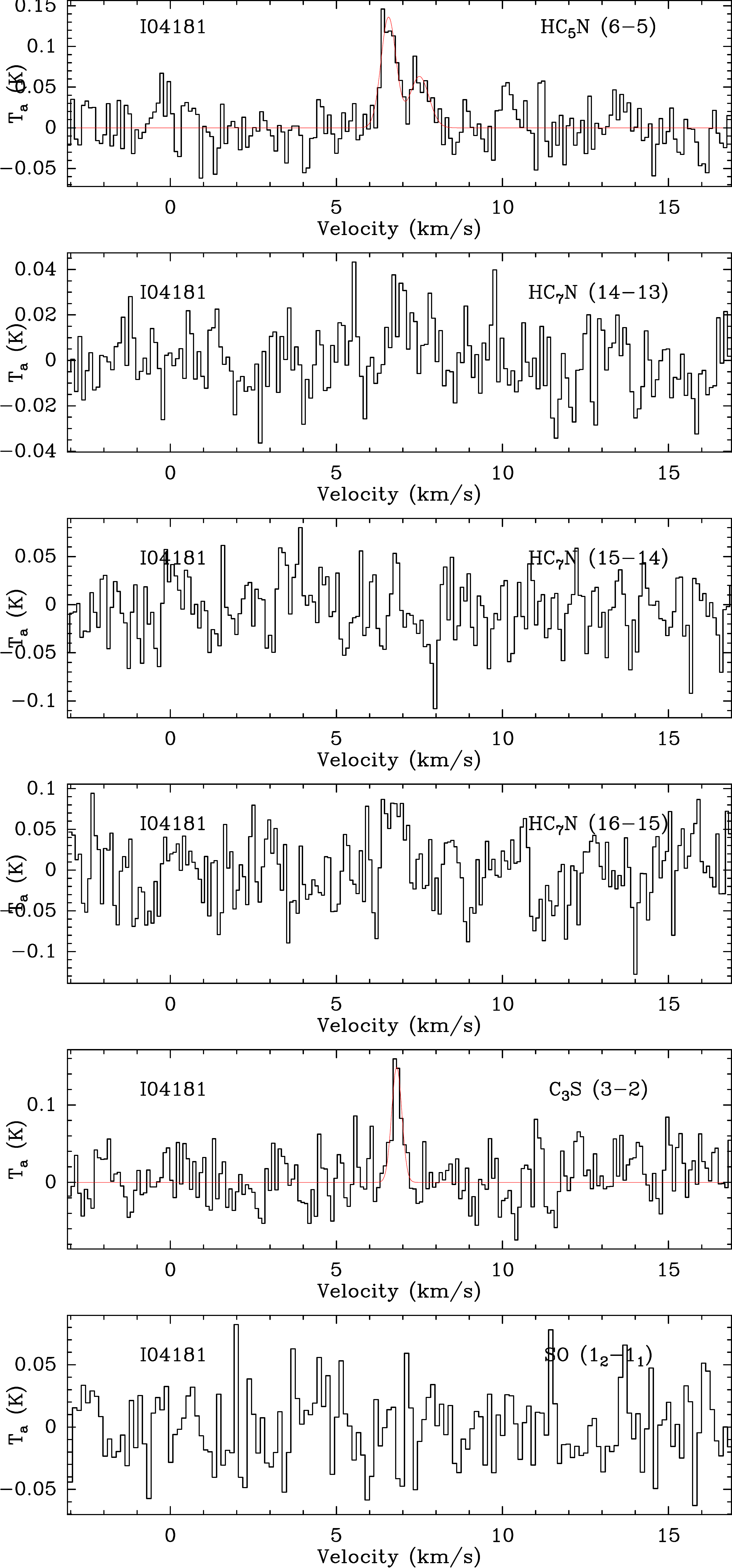}
  \includegraphics[width=0.24\linewidth]{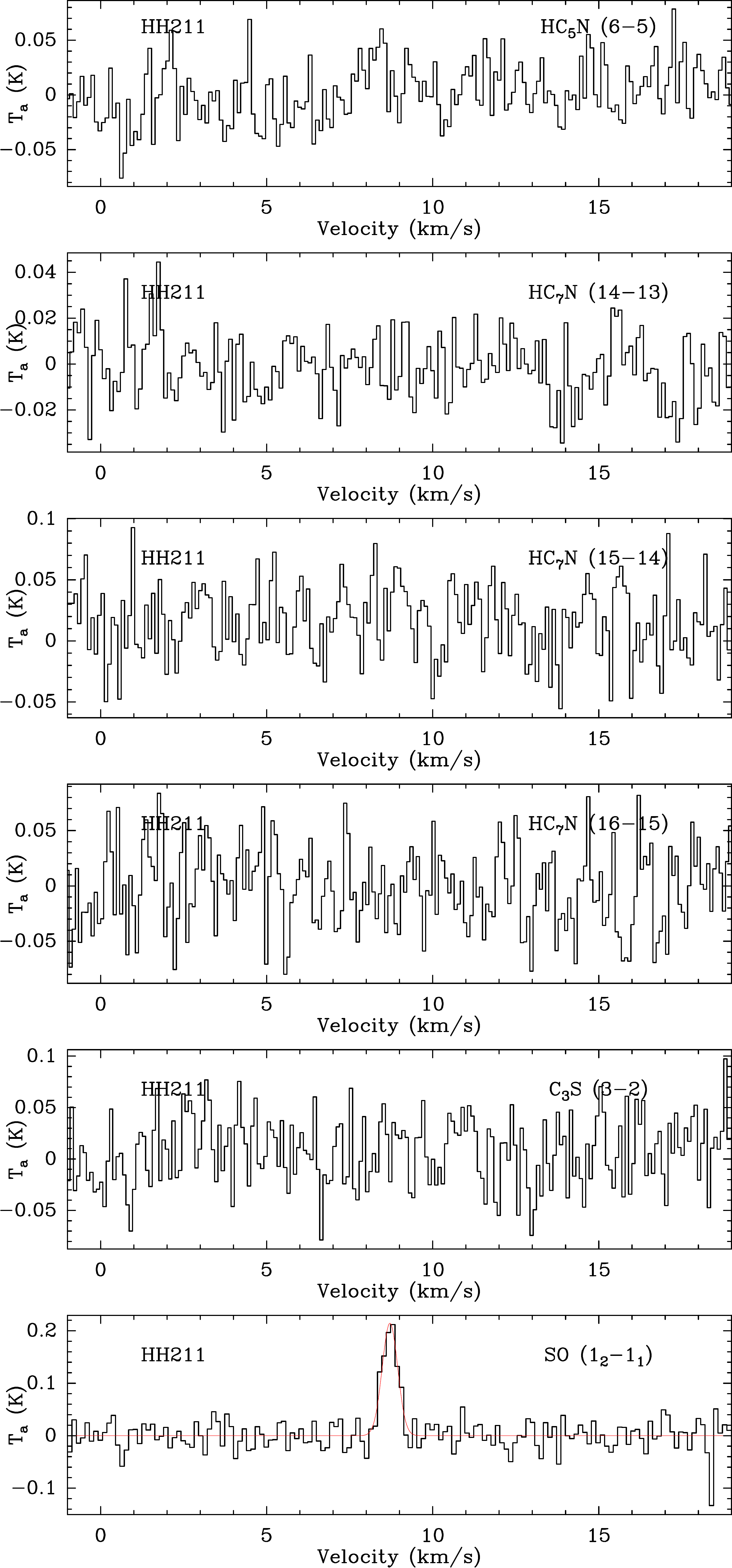}
  \includegraphics[width=0.24\linewidth]{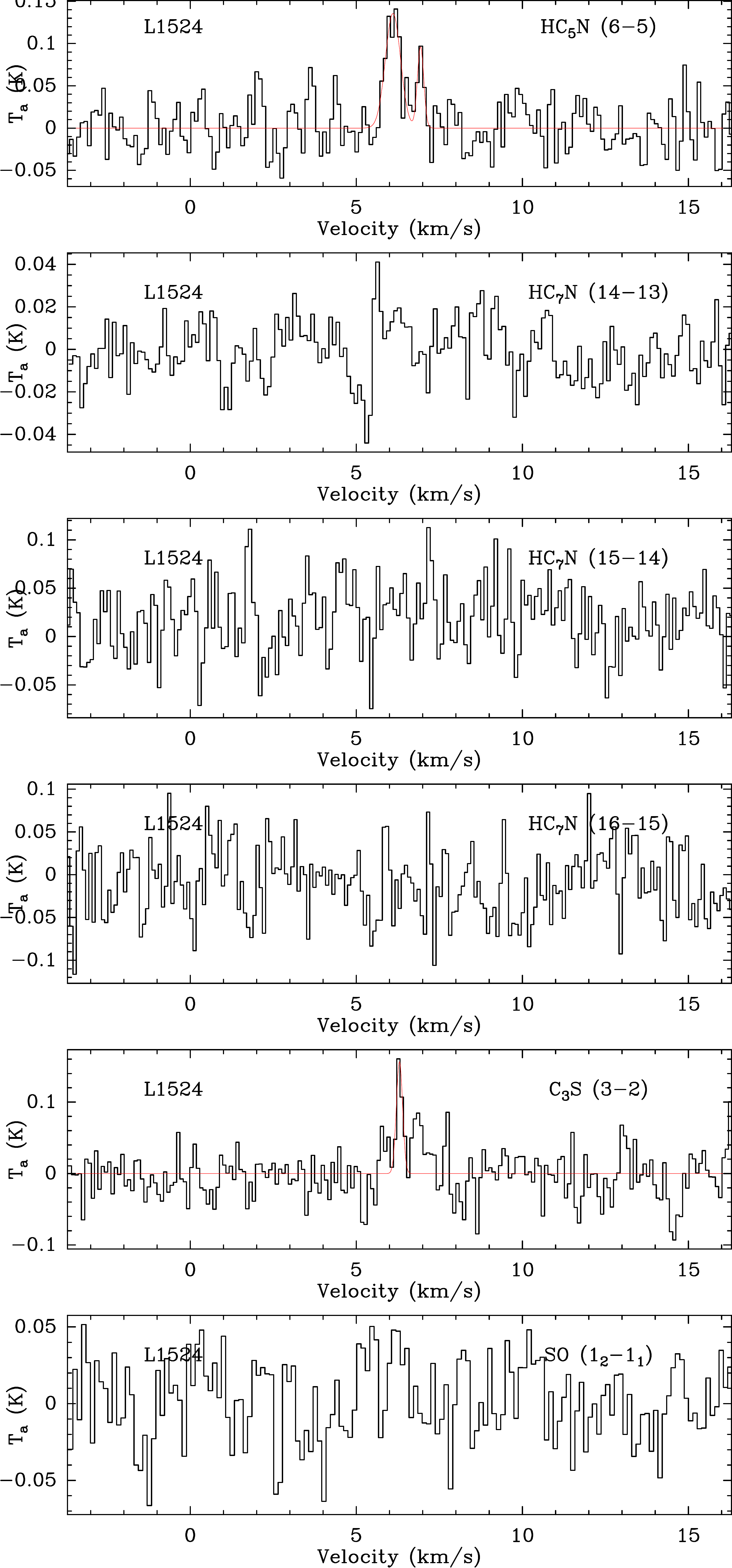}
  \includegraphics[width=0.24\linewidth]{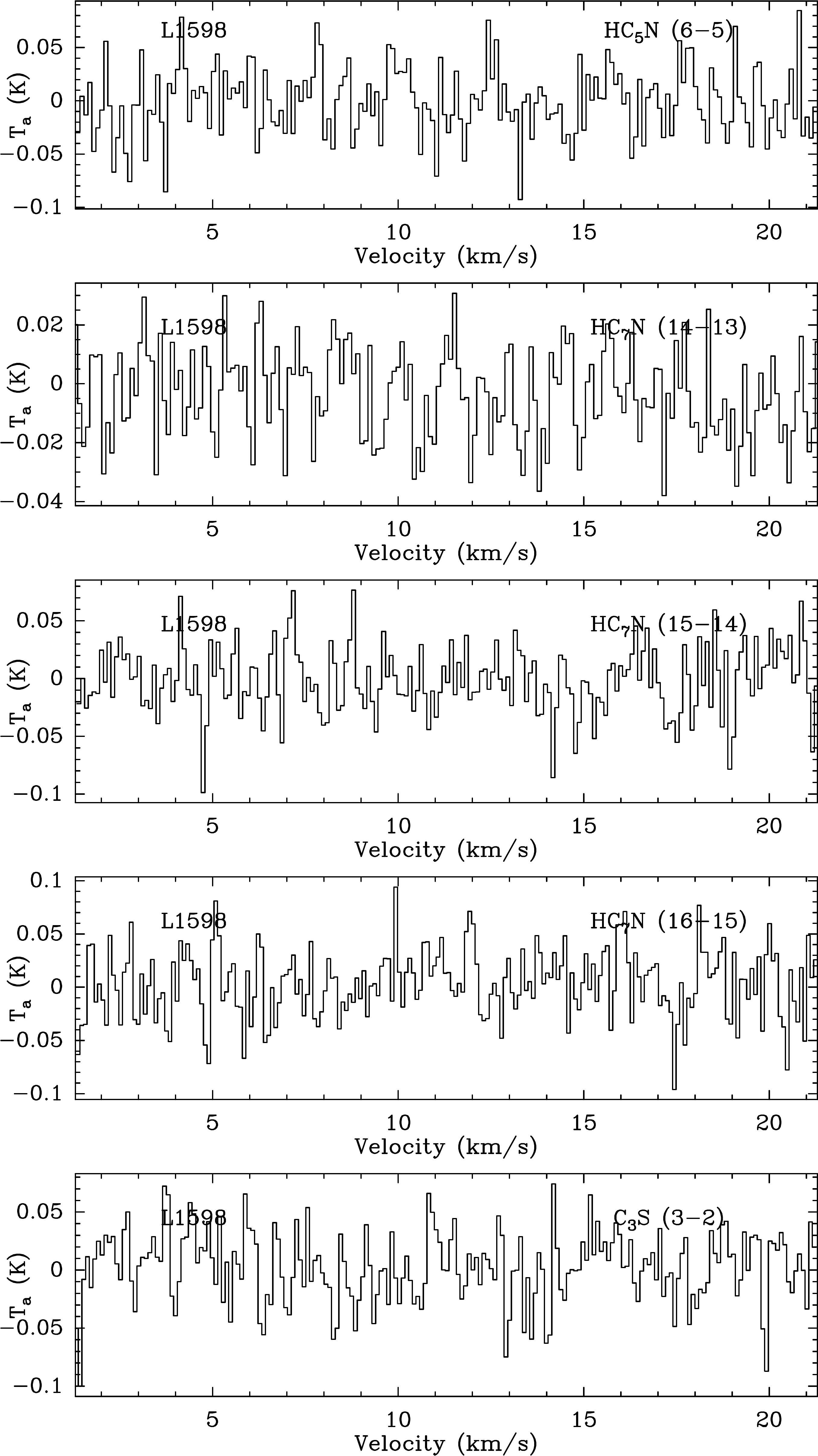}
  \caption{Spectra observed by TMRT. The source name and the transition are labeled at the upper left and upper right of each spectrum, respectively. \label{HC5N}}
\end{figure*}

In the Ku band, HC$_3$N (J=2--1) is detected in I04181, L1524, and HH211. HC$_5$N (J=6--5), and C$_3$S (J=3--2) are only detected in I04181 and L1524. SO (J=1$_2$--1$_1$) is only detected in HH211. These detected spectral lines are presented in Figs. \ref{HC3N} and \ref{HC5N}.
From Fig. \ref{HC3N}, we can see that four HFS components of HC$_3$N (J=2--1) are well resolved for the source L1524, and three HFS components of HC$_3$N (J=2--1) are detected in I04181. Only the main component, namely, F=3--2 of HC$_3$N (J=2--1), is detected in HH211. Due to the low S/N, the HFS components of HC$_5$N cannot be resolved. All these transitions are non-detected in L1598. L1598 is thus regarded as a source lacking CCMs.

The spectral lines were fitted with Gaussian function. The line parameters, including the centroid velocity V$_{\rm LSR}$, the antenna temperature T$_{\rm a}$, the full width at half maximum ($\Delta$V), and integrated intensities ($\int$T$_{\rm a}$dV), were obtained and they are listed in Tables \ref{line-para-PMO} and \ref{line-para-TMRT}. 
The line width and antenna temperature of the two groups of N$_2$H$^+$ with blended HFS lines were not obtained. The velocity integrated intensities are listed in Table \ref{line-para-PMO}. The error of the velocity integrated intensity was calculated as 
$T_{rms}\times\sqrt{N_{channels}}\times\delta V_{res}$, where T$_{\rm rms}$ is the 1$\sigma$ noise, N$_{\rm channels}$ is the total number of line channels, and $\delta V_{\rm res}$ is the velocity resolution. 
HH211 has highest integrated intensities of HNC and N$_2$H$^+$, followed by L1598. The integrated intensities of CCM species in L1598 are the weakest.

The HFS components of CCH (N=1--0), N$_2$H$^+$ (J=1--0), and HC$_3$N (J=2--1) were fitted by the HFS fitting program in GILDAS/CLASS. 
The HFS parameters of the antenna temperature multiply optical depth (T$_{\rm a}*\tau$), centroid velocity (V$_{\rm LSR}$), the full width at half maximum ($\Delta$V), and the optical depth ($\tau$) are listed in Table \ref{hfs}. The error of optical depth obtained by HFS fitting is considerable.

\subsection{Excitation}
Based on the optical depth ($\tau$) obtained by hyperfine structure fitting, the excitation temperature (T$_{\rm ex}$) can be derived by:
\begin{eqnarray}
  &T_{ex} =& \frac{h\nu}{k} ln^{-1}\left\{\frac{h\nu/k}{\frac{T_r}{f(1-exp(-\tau))}+J(T_{bg})}+1\right\},
\end{eqnarray}
where
\begin{eqnarray}
&J(T) =& \frac{h\nu/k}{exp(\frac{h\nu}{kT})-1},
\end{eqnarray}
and f is the beam filling factor, which is assumed to be 1, $\nu$ is the rest frequency, T$_{bg}$ (2.73 K) is the temperature of the cosmic background radiation, and T$_{\rm r}$=T$_{\rm a}$/$\eta$ is the brightness temperature. The derived T$_{\rm ex}$ of CCH, N$_2$H$^+$, and HC$_3$N are listed in column (6) of Table \ref{hfs}. The T$_{\rm ex}$ of CCH toward L1598 cannot be derived, which is caused by the low S/N and only three HFS components were detected.
The T$_{\rm ex}$ of CCH ranges from 4.8 to 6.4 K and is slightly larger than the values of L1498 and CB246 \citep[$\sim$4 K,][]{2009A&A...505.1199P}. Although the HFS components of N$_2$H$^+$ are not well resolved, we also carried out the HFS fitting and derived the T$_{\rm ex}$ and obtained a result ranging from 3.4 to 6.1 K, which is similar to the typical value of 5 K \citep{2002ApJ...572..238C}. The T$_{\rm ex}$ of HC$_3$N for L1524 is 8.7$\pm$3.9 K. The high level of error may come from the high level of uncertainty for the optical depth obtained from the HFS fitting.

Due to the lack of observed transitions from various excitation levels, the excitation temperatures of HNC, c--C$_3$H$_2$, CCS, CCCS, SO, HC$_5$N, and C$_4$H cannot be reliably derived. 
Under the local thermodynamic equilibrium (LTE) assumption, we took dust temperatures as the excitation temperatures to calculate the column densities of these molecules. The dust temperature is derived via a spectral energy distribution (SED) fitting (see Fig. \ref{SED}). 

\subsection{Column density and abundance}
The column densities of CCH were derived with the RADEX code\footnote{https://personal.sron.nl/$\sim$vdtak/radex/} \citep{2019ApJ...884..167T}, which is a computer program for non-LTE analyses of molecular lines toward interstellar clouds that are assumed to be homogeneous \citep{2007A&A...468..627V}. A data file for CCH, containing the energy levels, transition frequencies, Einstein A coefficients, and collisional rate coefficients with H$_2$, is needed to run RADEX. The file is provided by LAMDA database\footnote{https://home.strw.leidenuniv.nl/$\sim$moldata/} \citep{2005A&A...432..369S}. The kinetic temperature was assumed as dust temperature for each source. The volume densities were estimated by $\frac{\mathrm{N_{H_2}}}{\mathrm{D}\ \mathrm{HPBW}}$, where N$_{\mathrm{H_2}}$ is the H$_2$ column density (see Fig. \ref{SED}), D is the distance from the sun, HPBW is the half power beam width of 53$''$. The volume densities of I04181, HH211, L1524, and L1598 are 4.6 $\times$ 10$^4$ cm$^{-3}$, 24 $\times$ 10$^4$ cm$^{-3}$, 18 $\times$ 10$^4$ cm$^{-3}$, and 1.5 $\times$ 10$^4$ cm$^{-3}$ repectively. The background temperature is 2.73 K. We minimized the ratios of modeled intensities by RADEX and observed intensities minus 1 ($\frac{\mathrm{M_{flux}}}{\mathrm{O_{flux}}}$--1) of each line. The column densities of CCH are listed in Table \ref{density}, and error was estimated as $\sqrt{\sum{(\frac{\mathrm{M_{flux}}}{\mathrm{O_{flux}}}-1)^2}}$.

The column density of other species was derived by \citep{2015PASP..127..266M}:
\begin{eqnarray}
  N =&& \frac{3k}{8\pi^3\nu} \frac{Q}{\Sigma S_{ij}\mu^2} \frac{J(T_{ex}) exp\left(\frac{E_u}{kT_{ex}}\right) }{J(T_{ex})-J(T_{bg})} \int T_{r}dV,
\end{eqnarray}
where the permanent dipole moment, $\mu$, line strength, S$_{\rm ij}$, and upper level energy, E$_{\rm u}$, were adopted from the Cologne Database for Molecular Spectroscopy\footnote{http://www.astro.uni-koeln.de/cdms/} and listed in Table \ref{transitions}. The partition function $Q$ was estimated as $\frac{kT_{ex}}{hB}e^{\frac{hB}{3kT_{ex}}}$ \citep{2015PASP..127..266M}.
The column densities are listed in Table \ref{density}.

The large difference in column densities is evident via the use of HC$_3$N (J=2--1) and (J=10--9) transitions, as presented in Table \ref{density}. The effective excitation densities\footnote{Critical density has been widely used as a representative value for the density of the region traced by a certain transition. However, this is misleading and should only be used when no better alternative exists. As a case in point, widespread CCH has been observed in a low density region (n(H$_2$)  $\sim$10$^3$ cm$^{-3}$) when the corresponding critical density can be more than one order of magnitude higher. A better indicator of the density, namely, effective excitation density, has been proposed by \citet{Shirley_2015}} of HC$_3$N (J=2--1) and (J=10--9) are 9.7$\times$10$^2$ cm$^{-3}$ and 1.6$\times$10$^5$ cm$^{-3}$, respectively, when assuming the kinetic temperature to be 10 K \citep{Shirley_2015}. The higher-J HC$_3$N (J=10--9) is a good tracer of dense gas \citep{2020MNRAS.496.2790L}. The lower-J HC$_3$N (J=2--1) has a lower critical density than HC$_3$N (J=10--9) and thus it can trace more externally extended gas.
Moreover, higher-J HC$_3$N traces dense regions with sizes that are smaller than the beam size. Thus, the column density of HC$_3$N (J=10--9) can be underestimated. 
On the other hand, the underestimated column densities derived by the transitions with high upper-level energy (E$_{up}$/k$_B$) may be due to overestimated excitation temperature.
With the H$_2$ surface densities generated by the released Herschel data ( Fig. \ref{SED}), the abundances of these molecules were derived and listed in Table \ref{abundance}.

\subsection{Emission regions}
The integrated intensity maps of HNC (J=1--0), c--C$_3$H$_2$ (J=2$_{1,2}$--1$_{0,1}$), C$_4$H, and HC$_3$N (J=10--9) were made for the four sources, except the integrated intensity maps of C$_4$H and HC$_3$N (J=10--9) in L1598 due to the low S/N (see Figs. \ref{inte-map} and \ref{inte-map2}). As presented in Table \ref{line-para-PMO}, the integrated intensities of C$_4$H (J=17/2--15/2) and C$_4$H (J=19/2--17/2) in the four sources are all very weak, with values smaller than 0.1 K km s$^{-1}$. To better display the distribution of C$_4$H with higher S/N, the integrated intensities of these two transitions are added up and shown in Fig. \ref{inte-map2}. 
The first moment maps of HNC (J=1--0) emission for the four sources are seen in Fig. \ref{moment}. 

\subsubsection{IRAS 04181+2655}
\begin{figure*}
  \centering
  \subfigure[I04181]{
    \begin{minipage}[t]{0.48\linewidth}
    \includegraphics[width=1\linewidth]{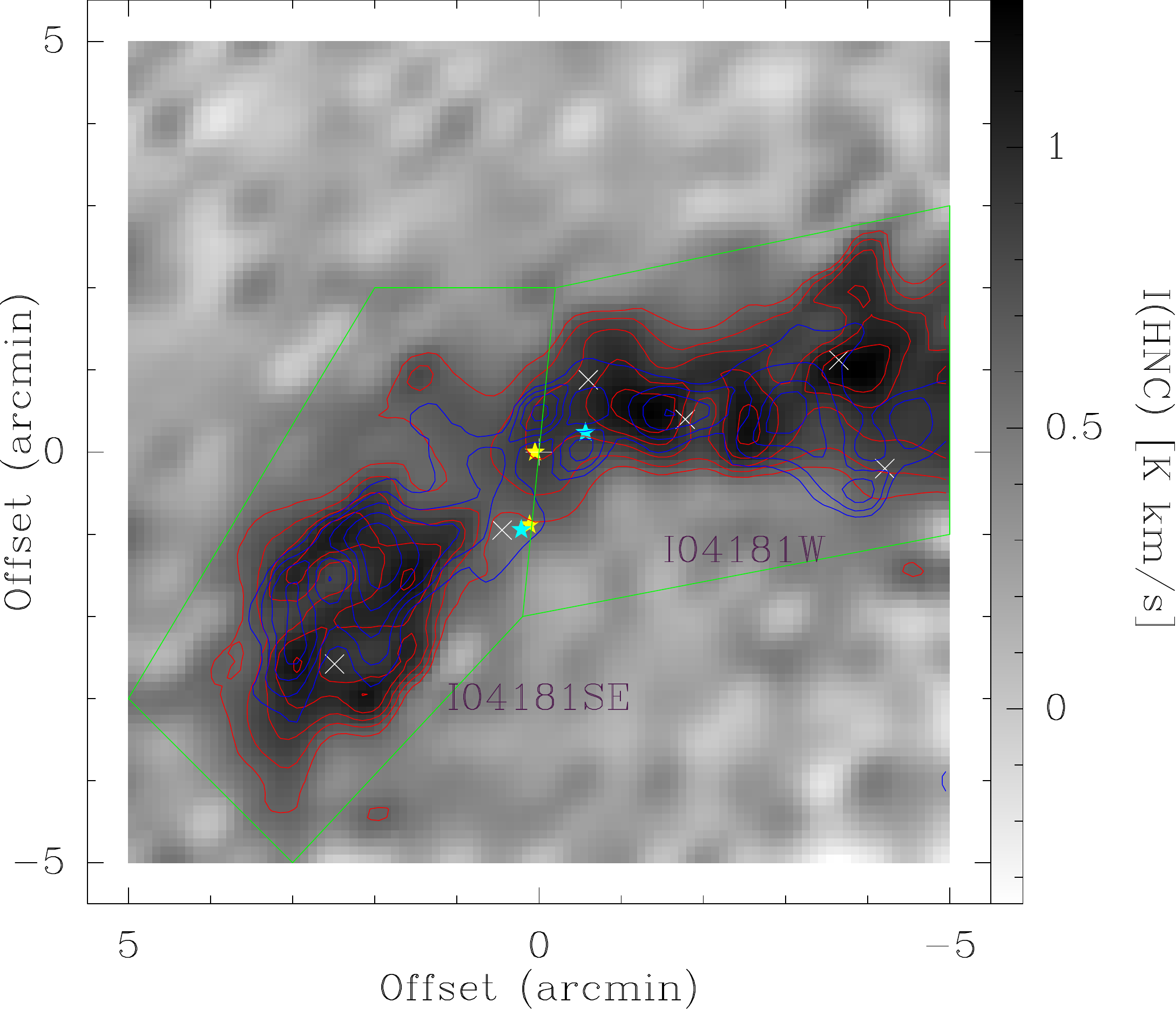}
    \end{minipage}}
\subfigure[HH211]{
  \begin{minipage}[t]{0.48\linewidth}
  \includegraphics[width=1\linewidth]{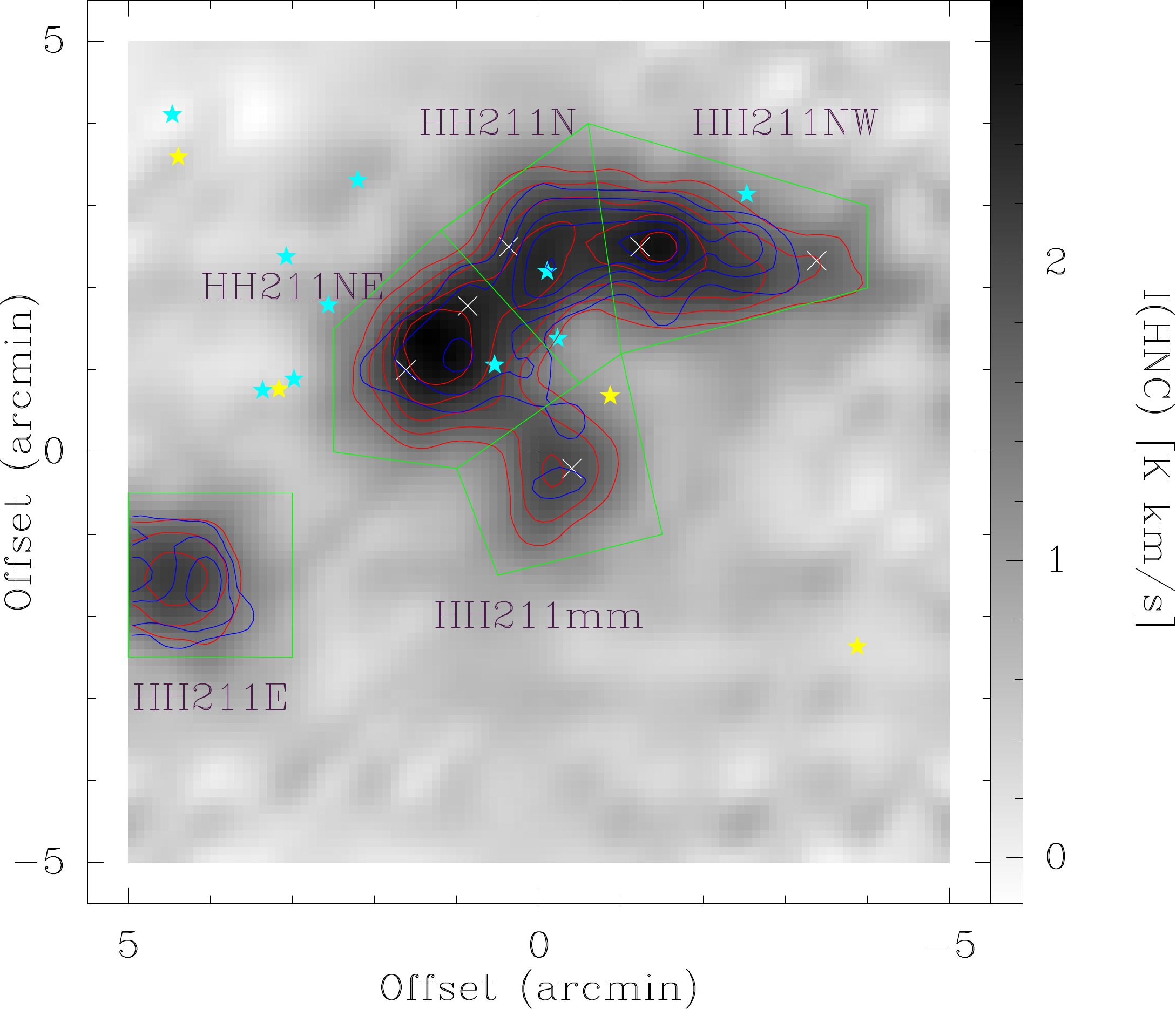}\\
\end{minipage}}
\subfigure[L1524]{
\begin{minipage}[t]{0.48\linewidth}
  \includegraphics[width=1\linewidth]{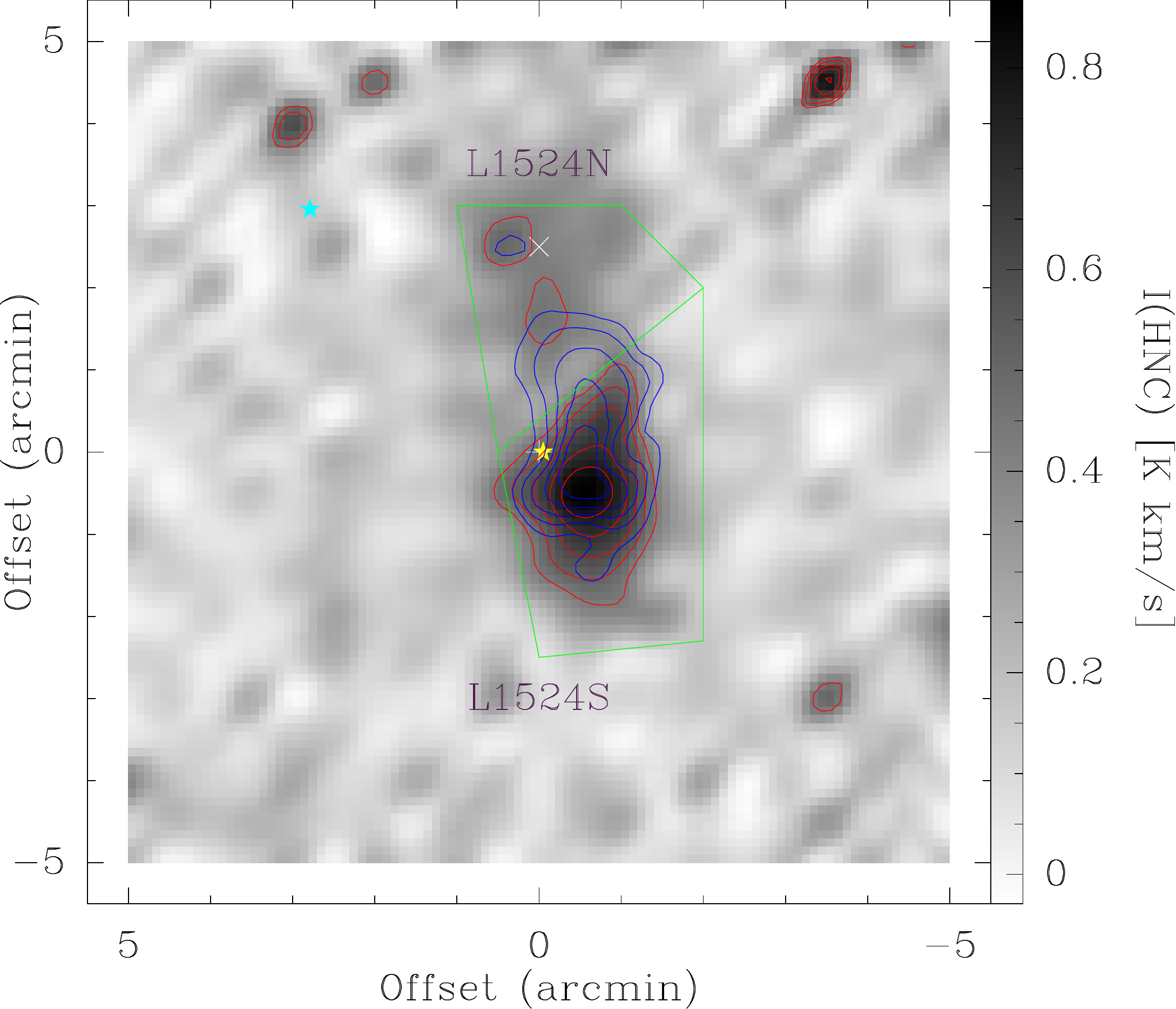}
\end{minipage}}
\subfigure[L1598]{
\begin{minipage}[t]{0.48\linewidth}
  \includegraphics[width=1\linewidth]{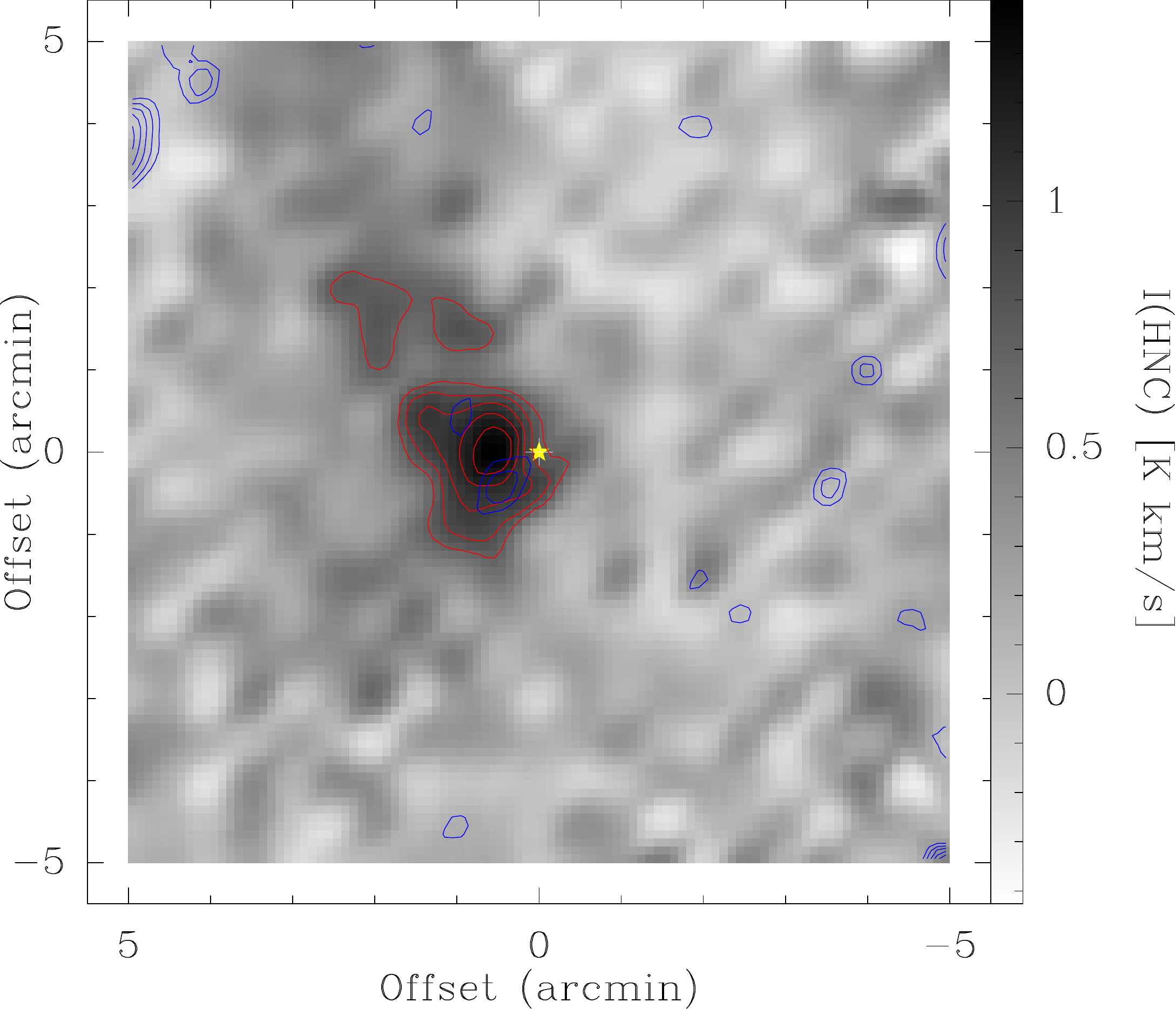}
\end{minipage}}
  \caption{Velocity-integrated intensity maps of HNC and c--C$_3$H$_2$. The integrated intensities of HNC and c--C$_3$H$_2$ are represented as red and blue contours, respectively, stepped from 50\% to 90\% by 10\% of the peak value. The greyscale shows the integrated intensity of HNC. The white cross ``+'' and ``x'' are marked as the source coordinates and the peak positions of ammonia emissions obtained from \citet{1989ApJ...341..208A,1987A&A...173..324B,2015ApJ...805..185S}. The yellow star and cyan star are marked as IRAS sources and the protostars presented by \citet{1988SSSC..C......0H,2016MNRAS.458.3479M}, respectively. \label{inte-map}}
\end{figure*}

\begin{figure*}
  \subfigure[I04181]{
    \begin{minipage}[t]{0.48\linewidth}
    \includegraphics[width=1\linewidth]{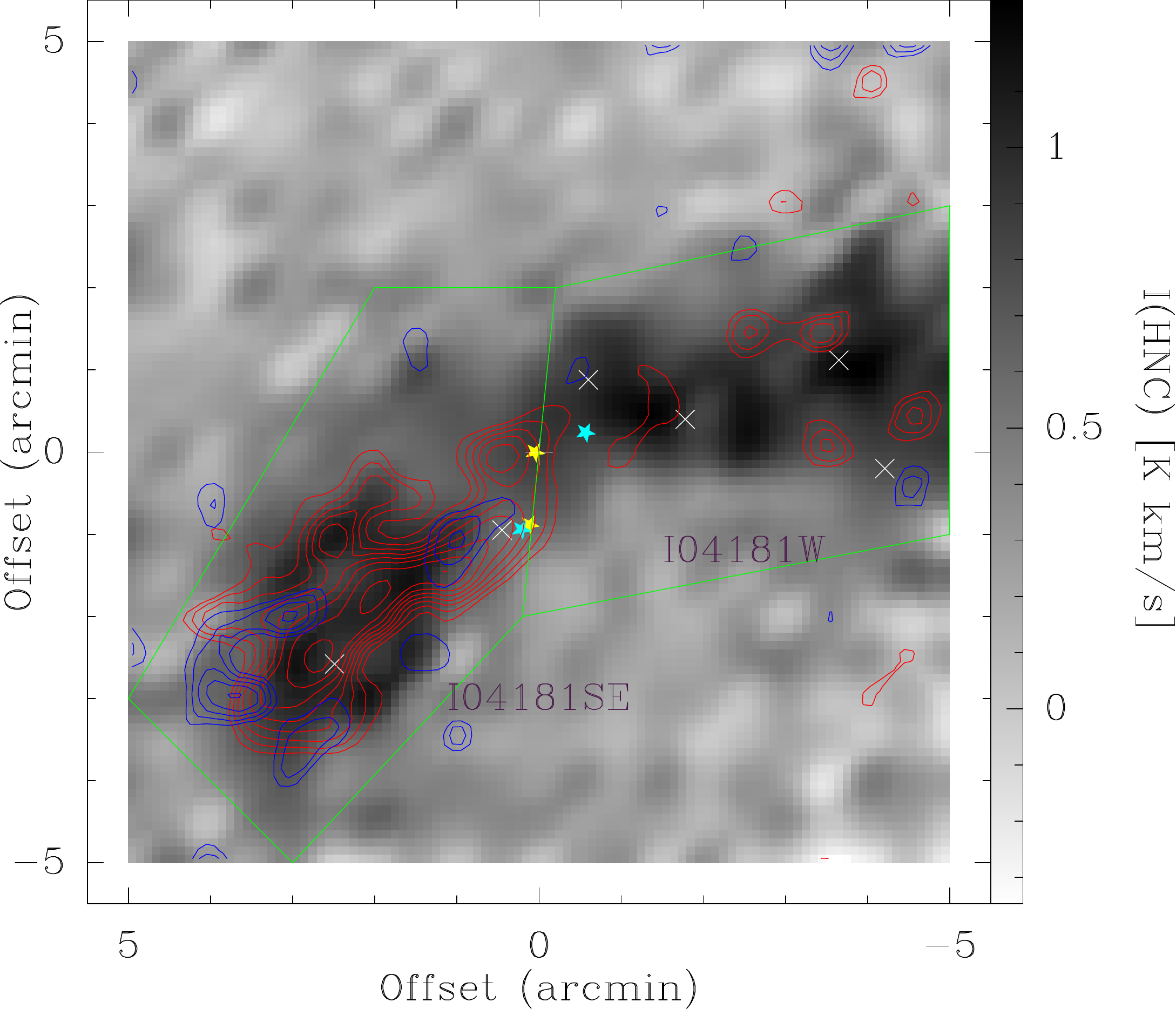}
    \end{minipage}}
\subfigure[HH211]{
  \begin{minipage}[t]{0.48\linewidth}
  \includegraphics[width=1\linewidth]{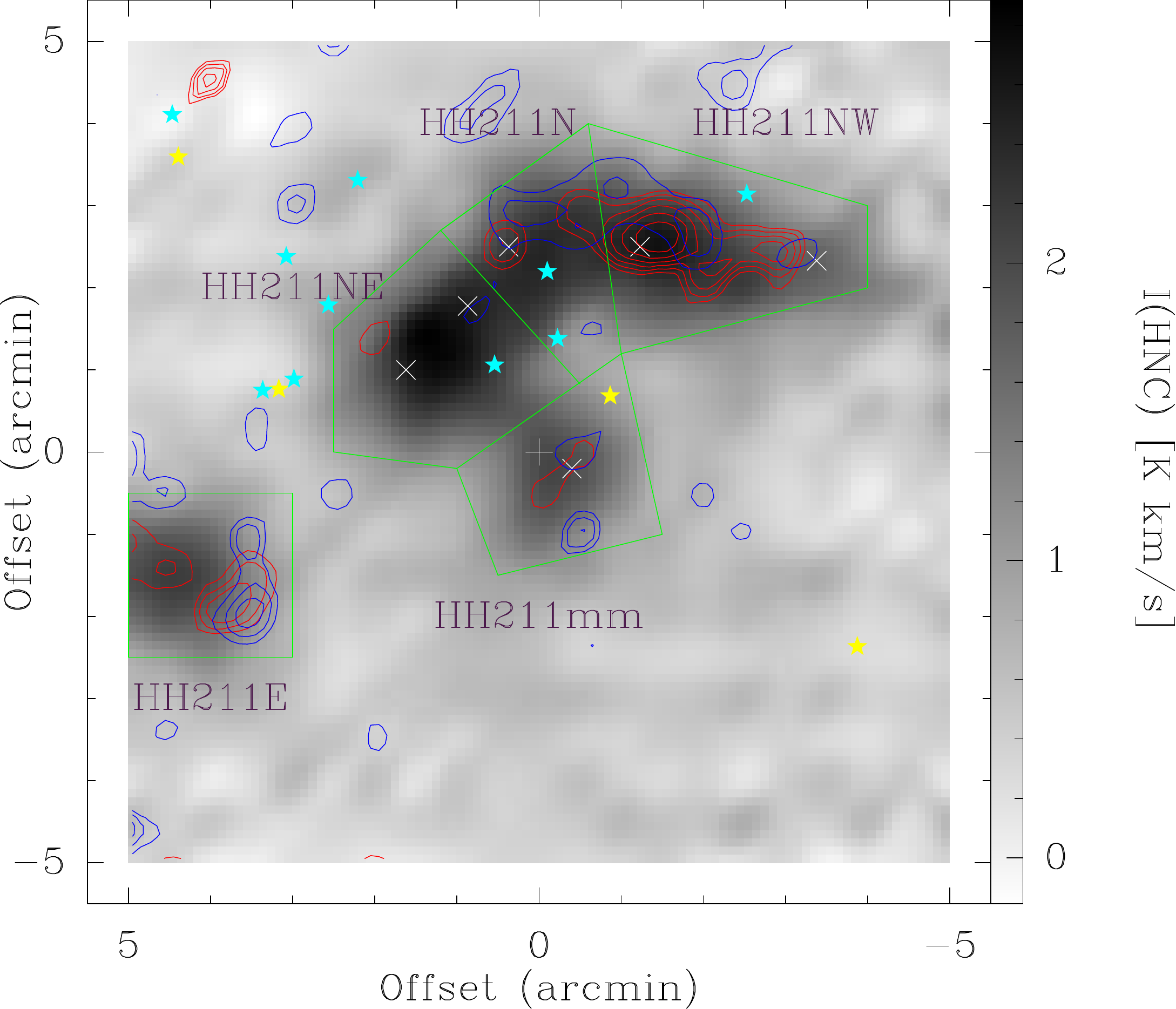}\\
\end{minipage}}
\subfigure[L1524]{
\begin{minipage}[t]{0.48\linewidth}
  \includegraphics[width=1\linewidth]{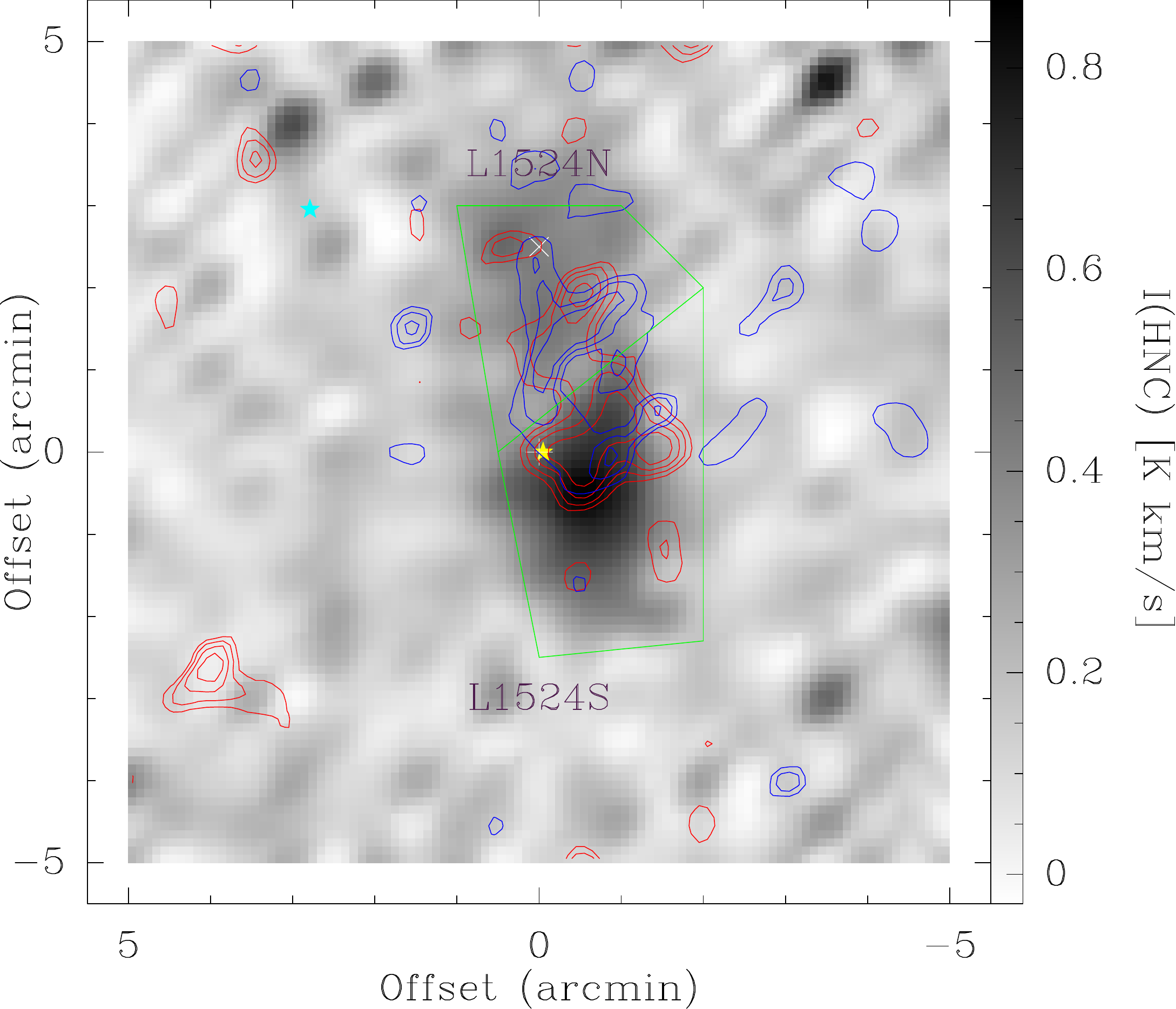}
\end{minipage}}
  \caption{Velocity-integrated intensity maps of HC$_3$N and C$_4$H. The marker sets and gray scale are the same as Fig. \ref{inte-map}. The integrated intensities of HC$_3$N and C$_4$H are represented as red and blue contours, respectively, stepped from 5$\sigma$ to the maximum by 1$\sigma$. The 1$\sigma$ values of HC$_3$N are 0.036 K km s$^{-1}$ for I04181 and 0.03 K km s$^{-1}$ for HH211 and L1524. The 1$\sigma$ value of C$_4$H is 0.05 K km s$^{-1}$.\label{inte-map2}}
\end{figure*}

According to the HNC and HC$_3$N image maps (see Figs. \ref{inte-map} and \ref{inte-map2}), we divided I04181 into I04181SE and I04181W. The ring-like structures are shown in the emission regions of HNC and c--C$_3$H$_2$ toward I04181SE, which is presented in Fig. \ref{inte-map}. C$_4$H emission seems to surround the HC$_3$N emission found in I04181SE. There is an obvious velocity gradient between I04181SE and I04181W ranging from 6.7 km s$^{-1}$ to 7.2 km s$^{-1}$.

\subsubsection{HH211}
According to HNC emissions, the HH211 region was divided into HH211mm, HH211E, HH211NE, HH211N, and HH211NW. HH211mm harbors a Class 0 protostar. The Class 0 protostar IC348mm is located in HH211N \citep{2013ApJ...768..110C}.
It seems that the velocity gradient transition from east to west occurred in HH211mm, HH211NE, and HH211N. HH211E has the largest LSR velocity among these cores and thus may be far away from the other gas regions.

\subsubsection{L1524}
\begin{figure*}
  \centering
  \subfigure[I04181]{
    \begin{minipage}[t]{0.45\linewidth}
    \includegraphics[width=1\linewidth]{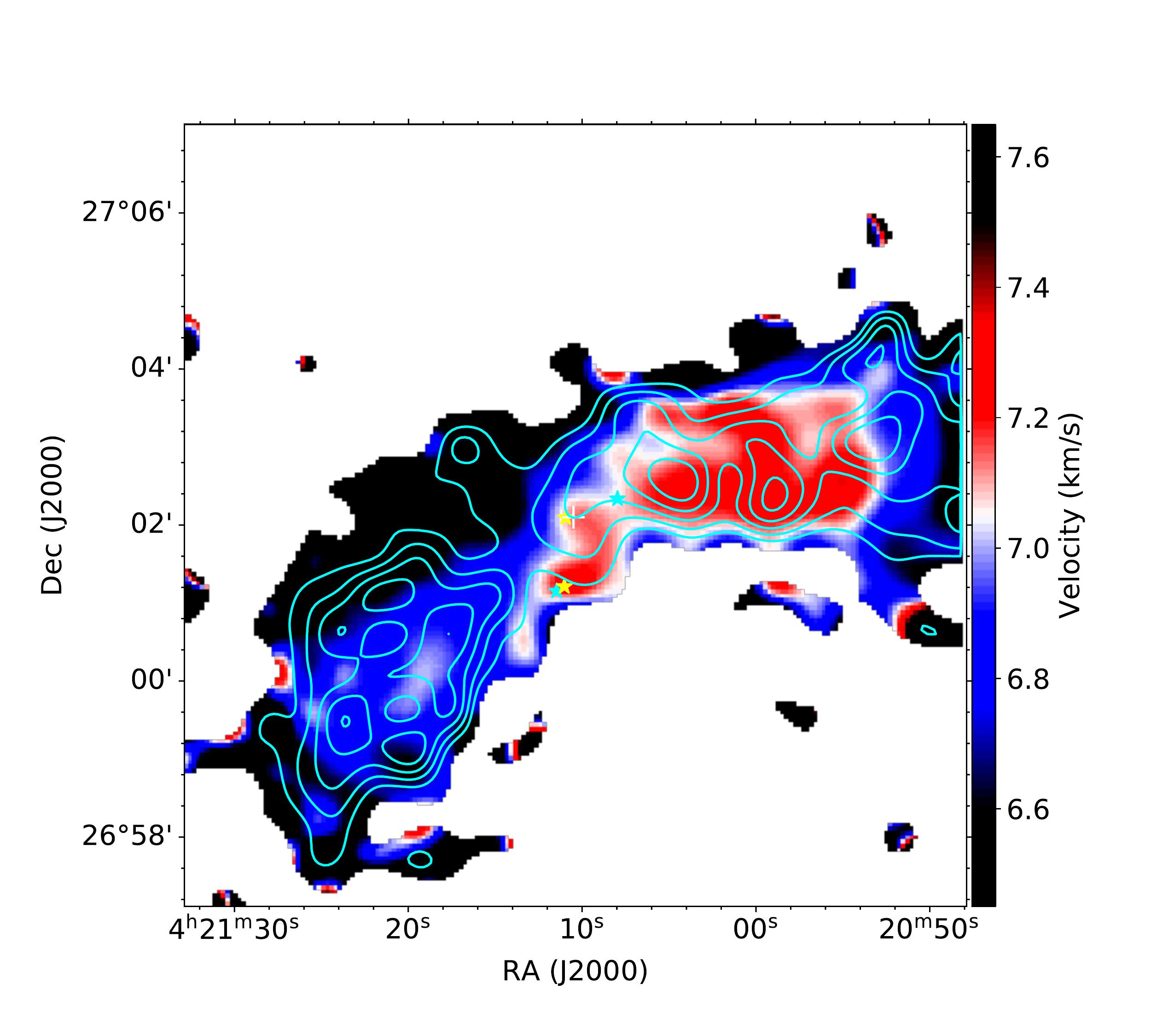}
    \end{minipage}}
\subfigure[HH211]{
  \begin{minipage}[t]{0.45\linewidth}
  \includegraphics[width=1\linewidth]{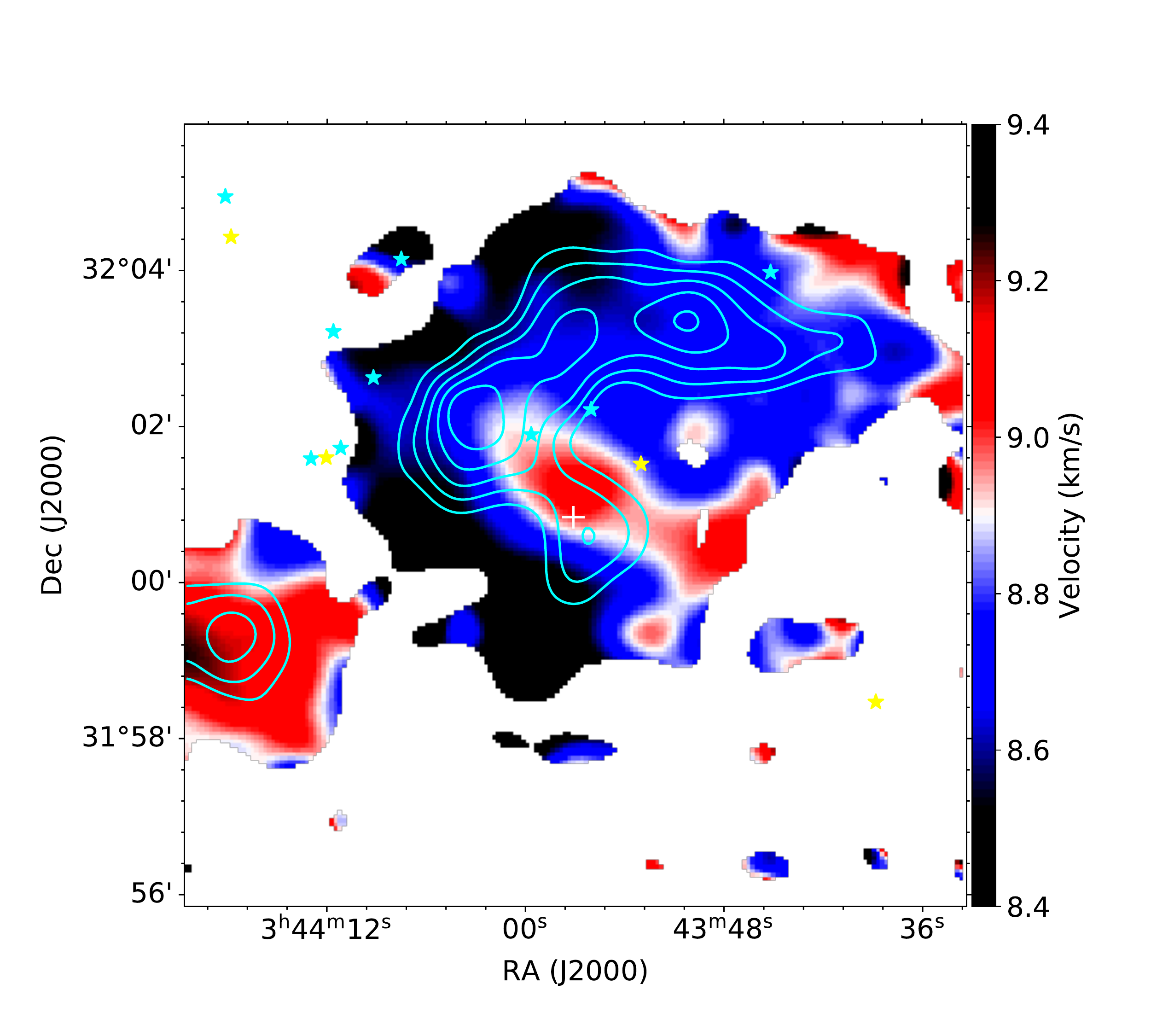}\\
\end{minipage}}
\subfigure[L1524]{
\begin{minipage}[t]{0.45\linewidth}
  \includegraphics[width=1\linewidth]{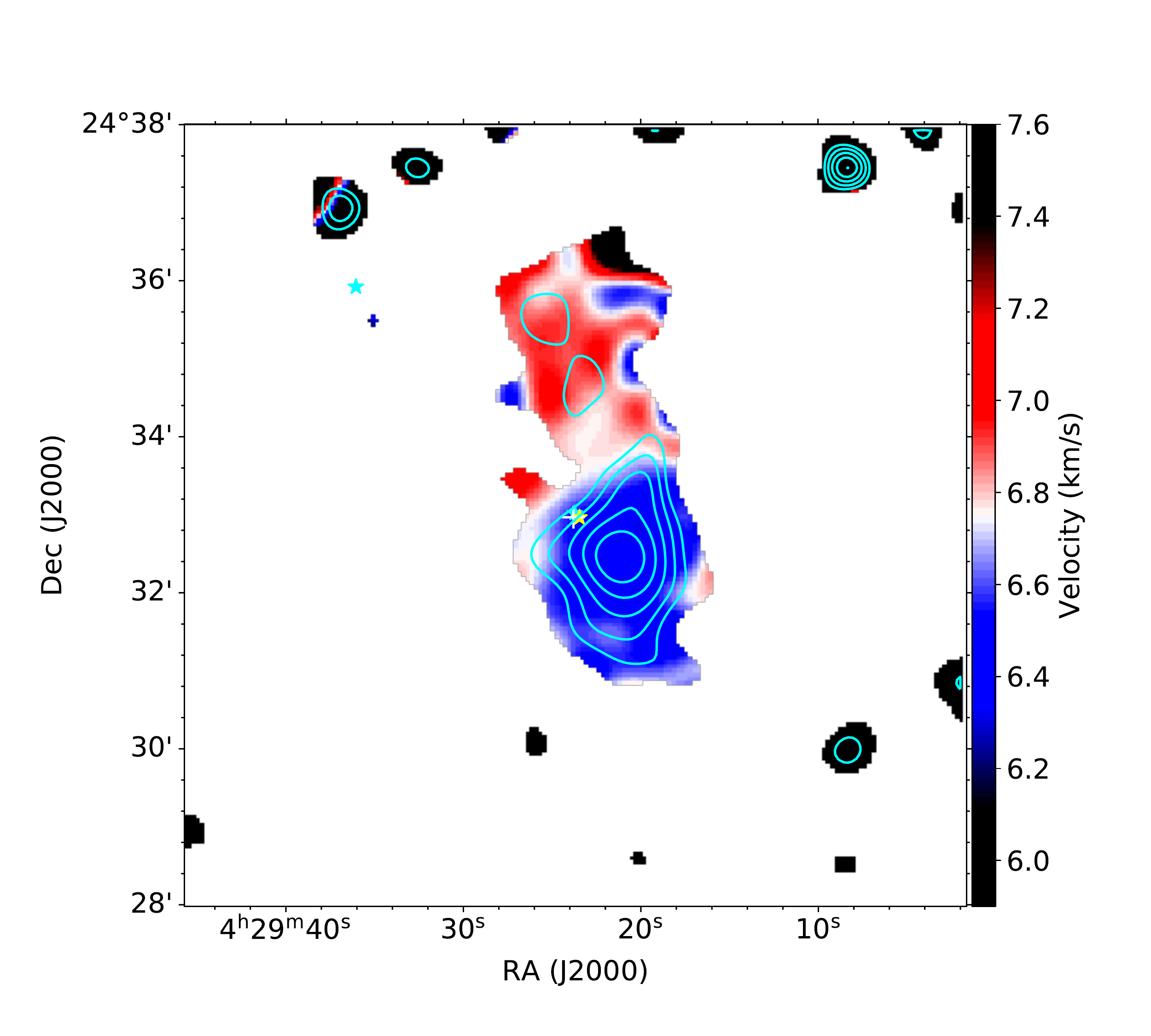}
\end{minipage}}
\subfigure[L1598]{
\begin{minipage}[t]{0.45\linewidth}
  \includegraphics[width=1\linewidth]{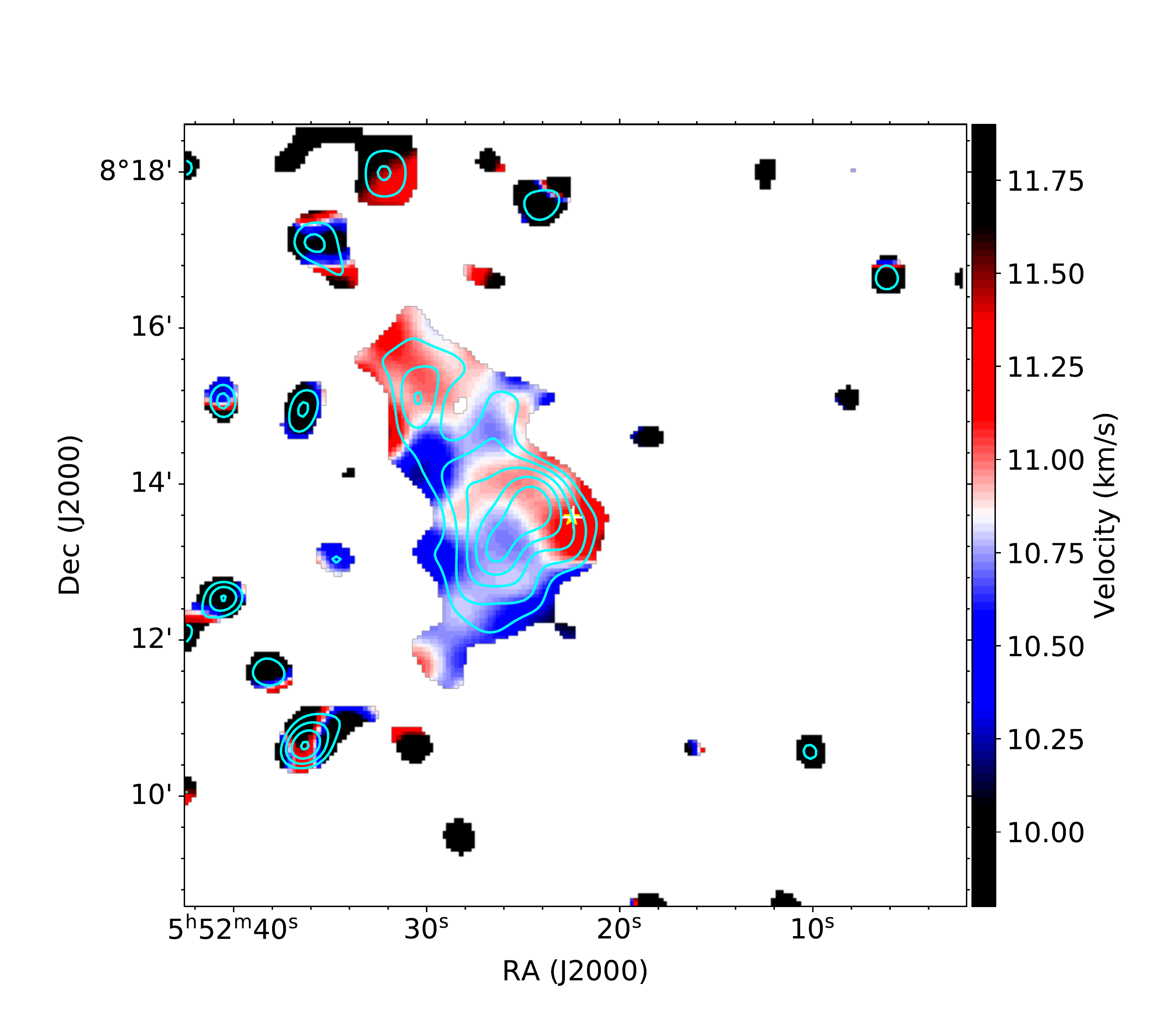}
\end{minipage}}
  \caption{Color image of the first moment maps of HNC (J=1--0) emission. Contours represent the integrated intensity maps of HNC with contour levels stepped from 50\% to 90\% by 10\% of the peak value. The marker sets are the same as Fig. \ref{inte-map} except the peak positions of ammonia emissions which are not marked. \label{moment}}
\end{figure*}

The weakest HNC emission is found in L1524 with a peak value of $\sim$0.8 K km s$^{-1}$ (see Fig. \ref{inte-map}). The ammonia condensation peaks at 2.5$'$ north and distributes in the northern regions \citep{1989ApJ...341..208A}. We divided the L1524 region into L1524N and L1524S. L1524S have stronger HC$_3$N emission but weaker C$_4$H emission than L1524N (see Fig. \ref{inte-map2}). In Fig. \ref{moment}, the difference of LSR velocity between L1524N and L1524S is $\sim$0.5 km s$^{-1}$.

\subsubsection{L1598}
HNC emissions are very strong in L1598 (see Table ref{line-para-PMO}), but the emissions of HC$_3$N and C$_4$H are too weak to generate the integrated intensity maps.

As shown in Fig. \ref{inte-map}, the HNC gas cores always deviate from the protostars, which can be seen in I04181, L1524, and L1598. In the HH211mm region, HNC emission is gathering toward Class 0 protostar HH211. Relative to the HNC emission region, the c--C$_3$H$_2$ (J=2$_{1,2}$--1$_{0,1}$) emissions are abundant around the protostar in I04181 and L1524.

\section{Discussion \label{discuss}}
\subsection{Chemical status}
\subsubsection{The molecular pair of CCH and N$_2$H$^+$\label{CCH_vs_N2H+}}
In the early stage of dark clouds, CCH is the most abundant hydrocarbon \citep{2018ApJ...856..151L} with an extended distribution \citep{2008ApJ...675L..33B,2017ApJ...836..194P}, while N$_2$H$^+$ formation is impeded as a result of CO gas possibly reacting with its precursor H$_3^+$ and which may directly destroy N$_2$H$^+$ through the reaction $\mathrm{N}_2\mathrm{H}^++\mathrm{CO}\rightarrow\mathrm{HCO}^++\mathrm{N}_2$ \citep{2002ApJ...570L.101B}. 
In the latter stage of cores, CCH is oxidized to form other species such as CO, OH, and H$_2$O in the dense center regions \citep[e.g.,][]{2008ApJ...675L..33B,2014A&A...562A...3M,2016A&A...592A..21F}, while N$_2$H$^+$ shows centrally peaked emission in the dense region where CO depletion has occurred \citep{2002ApJ...570L.101B}. The two kinds of molecules can be regarded as good evolution tracers for the Planck Galactic cold clump cores \citep{2019A&A...622A..32L} as the abundance ratio of x[CCH]/x[N$_2$H$^+$] decreases with evolution.

\begin{figure*}
  \centering
  \includegraphics[width=0.45\linewidth]{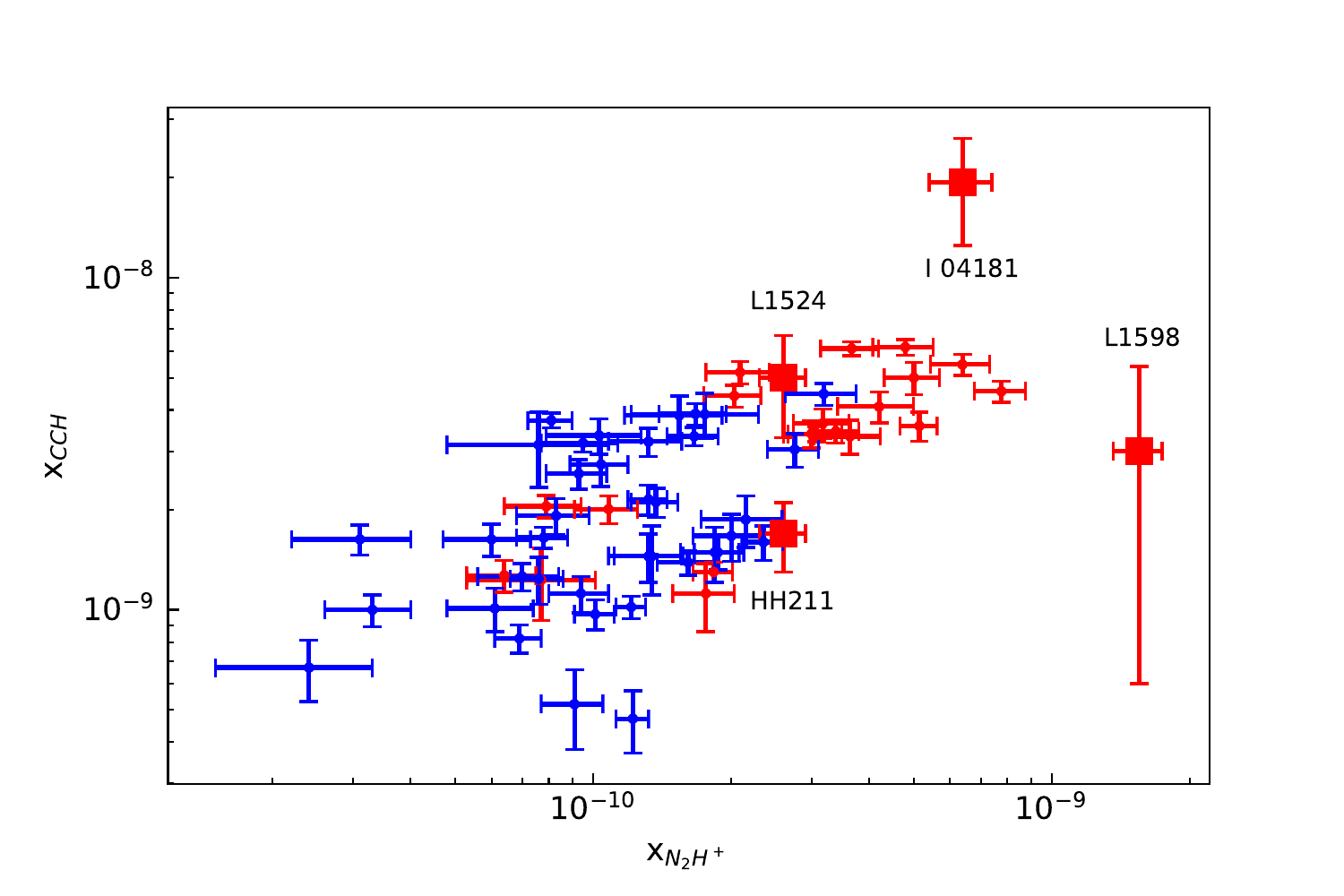}
  \includegraphics[width=0.45\linewidth]{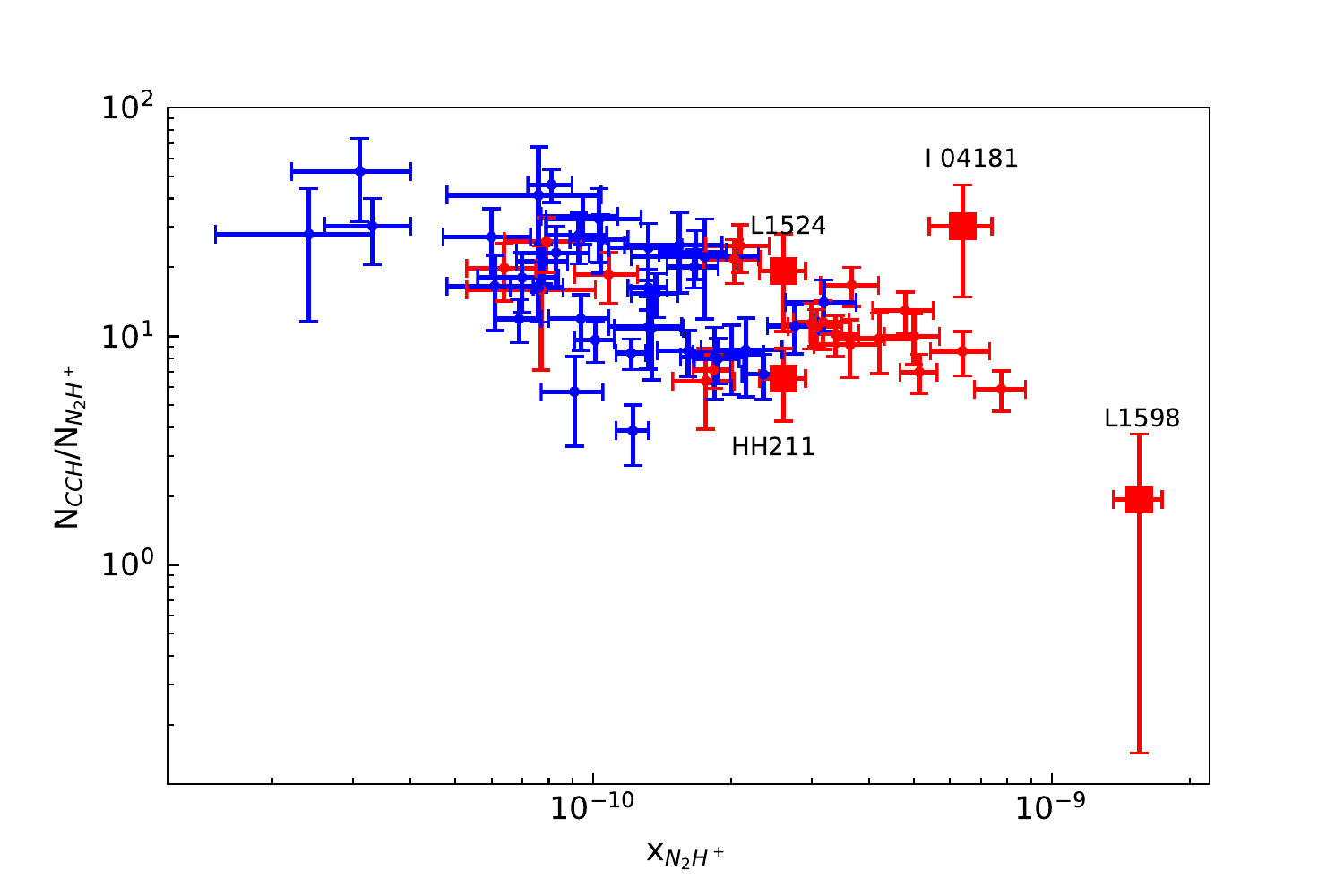}
  \caption{Relationships among x(N$_2$H$^+$), x(CCH), and N(CCH)/N(N$_2$H$^+$). Left: x(N$_2$H$^+$) v.s. x(CCH). Right: N(CCH)/N(N$_2$H$^+$) v.s. x(N$_2$H$^+$). Sources of I04181, HH211, L1524 and L1598 are marked as red squared data points. The circled data points are obtained from \citet{2019A&A...622A..32L}. The circled data points are marked in red or blue according to whether or not they harbor an IRAS source. \label{N2H_vs_CCH}}
\end{figure*}

The relationships among the abundances of CCH and N$_2$H$^+$ and their ratios are plotted in Fig. \ref{N2H_vs_CCH}, where our four sources are marked as red squares and the circled data points are obtained from \citet{2019A&A...622A..32L}. The red circled data points are marked as harboring the \emph{IRAS} source within 1 arcmin \citep{1988SSSC..C......0H} and, thus, the corresponding cores are believed to contain heating sources. These cores are considered to be in the late stage of star formation. The early-stage core is shown in blue. In the left panel of Fig. \ref{N2H_vs_CCH}, it is clear that most of the red data points are located in the top-right corner and most of the blue data points are located in the bottom-left corner. It seems that the abundances of CCH and N$_2$H$^+$ increase when going from starless cores to star-forming cores. Two facts may offer an explanation for the increasing CCH abundance:\ 1) CCH arises from dense gas with densities of n(H$_2$)$\sim$10$^4$-10$^5$ cm$^{-3}$ but not from the densest regions with n(H$_2$) > 10$^6$ cm$^{-3}$ \citep{1988A&A...195..257W}. In the four sources, the volume densities of H$_2$ range from 1.5$\times$10$^4$ to 2.4$\times$10$^5$ cm$^{-3}$ within a beam size of 53$''$;
2) the gas can be affected by star-forming feedback. Heating and radiation may lead to regenerated CCH.
As shown in the right panel of Fig. \ref{N2H_vs_CCH}, N[CCH]/N[N$_2$H$^+$] decreases from early stage cores to latter stage cores. For the four star-forming cores, more evolved cores of I04181 and L1524 have a value of N[CCH]/N[N$_2$H$^+$] that is larger than that of HH211. However, L1598, which is also an evolved core, has the lowest ratio. The ratio in our four protostars may
reflect source variety, or L1598 is a special source with absent CCH.

\subsubsection{Dense gas tracers of N$_2$H$^+$, C$_4$H, and HC$_3$N \label{N2H+_vs_HC3N}}
\begin{figure*}
  \centering
  \includegraphics[width=0.45\linewidth]{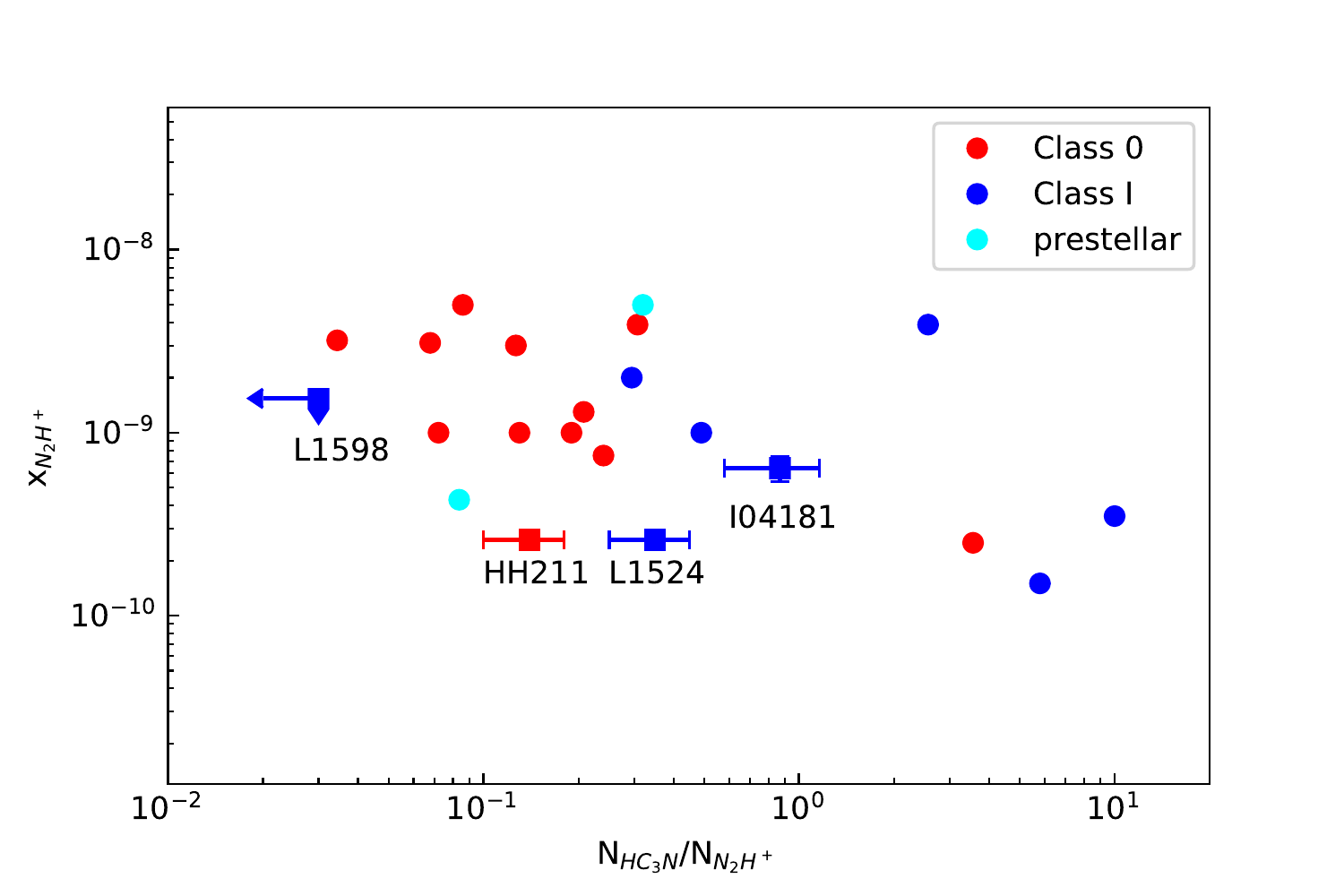}
  \includegraphics[width=0.45\linewidth]{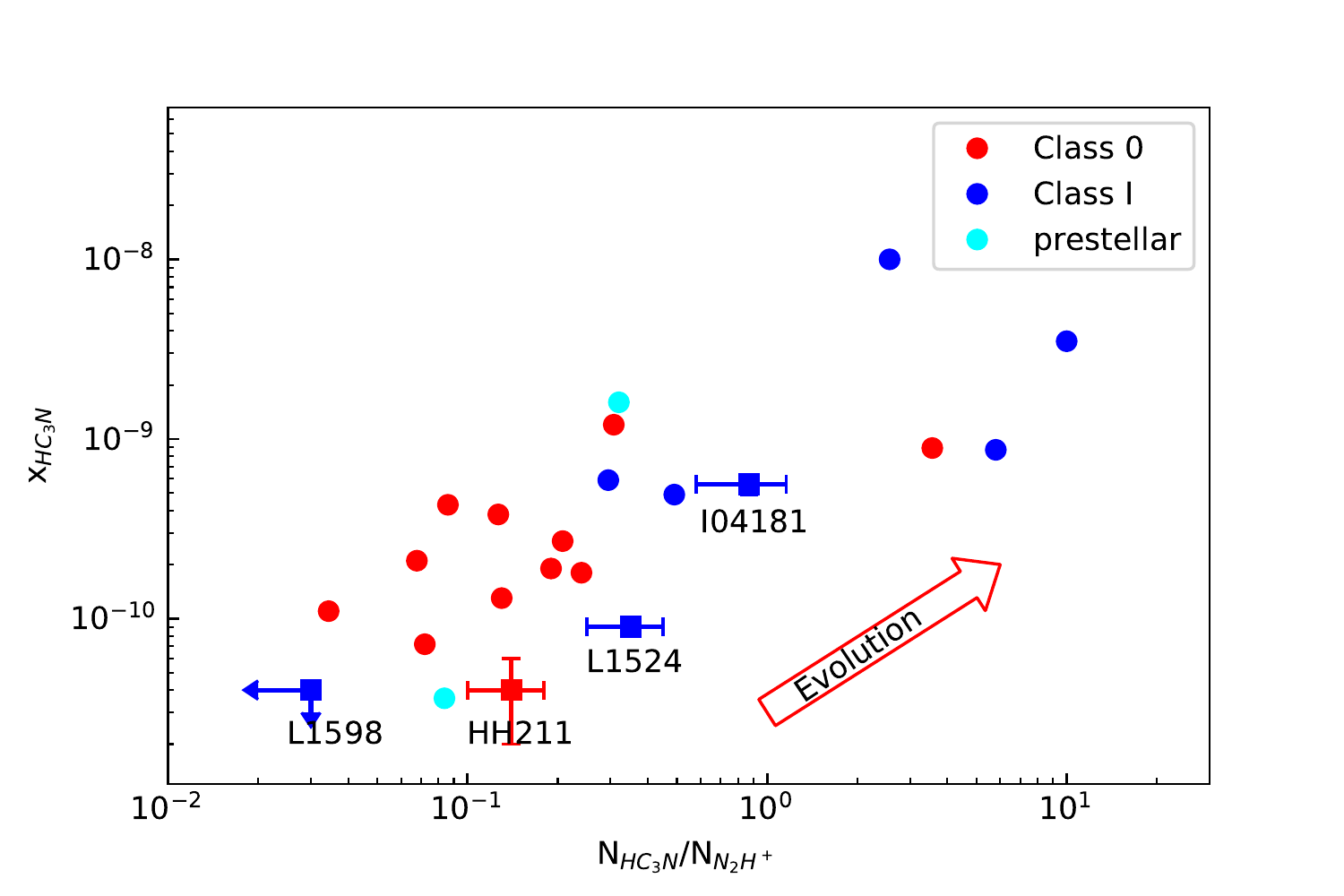}
  \caption{Relationships among x(N$_2$H$^+$), x(HC$_3$N), and N(N$_2$H$^+$)/N(HC$_3$N). Left: x(N$_2$H$^+$) v.s. N(N$_2$H$^+$)/N(HC$_3$N). Right: x(HC$_3$N) v.s. N(N$_2$H$^+$)/N(HC$_3$N). Sources of I04181, HH211, L1524, and L1598 are marked as squared data points.  Circled data points are obtained from \citet{2004A&A...416..603J}. The colors of cyan, red, and blue are marked as prestellar, Class 0, and Class I objects, respectively. \label{N2H_vs_HC3N}}
\end{figure*}

CCMs are formed from C and C$^+$ which are abundant in photo-dissociation regions \citep[PDR,][]{2005A&A...435..885P}. Thus, CCMs are often used as extended gas tracers \citep[and references therein]{2015ApJ...808..114J}. 
However, C$_4$H and HC$_3$N can be formed from CH$_4$ \citep{1984MNRAS.207..405M,2008ApJ...681.1385H} and C$_2$H$_2$ \citep{2009MNRAS.394..221C}, which are sublimated from grain mantles with dust temperatures of 25 K and 50 K, respectively \citep{1983A&A...122..171Y,1984MNRAS.207..405M}. Regarding the hot components of HC$_3$N, its vibrational excitation lines have been detected around massive young stellar objects \citep[e.g.,][]{2020ApJ...898...54T}. Thus, high J HC$_3$N are capable of tracing dense cores quite well \citep{2020MNRAS.496.2790L}.
The abundances of C$_4$H and HC$_3$N increase in high-mass protostellar objects (HMPOs) and WCCC sources \citep{2008ApJ...672..371S,2019ApJ...872..154T}. In these places, dust temperatures are higher than 30 K and can be up to 50 K \citep{2002ApJ...566..931S,2008ApJ...672..371S}. Recently, the molecular pair of HC$_3$N and N$_2$H$^+$ have been used as a chemical evolutionary indicator in massive star-forming regions \citep{2019ApJ...872..154T}.
We plotted the relationships among N(HC$_3$N)/N(N$_2$H$^+$), x(N$_2$H$^+$), and x(HC$_3$N) as shown in Fig. \ref{N2H_vs_HC3N} combined with the data from \citet{2004A&A...416..603J}. The error of the data was not given in \citet{2004A&A...416..603J}.
We can see that the N(HC$_3$N)/N(N$_2$H$^+$) ratio and the x(HC$_3$N) value tend to increase from Class 0 to Class I in Fig. \ref{N2H_vs_HC3N}, which is consistent with the trend found in massive star-forming regions \citep{2019ApJ...872..154T}. We summarized the statistical analyses in Table \ref{stt}. The mean values of N(HC$_3$N)/N(N$_2$H$^+$) ratio are 0.43 and 2.55 in Class 0 and Class I protostars, respectively. The mean values of x(HC$_3$N) are 3.4$\times$10$^{-10}$ and 20.2$\times$10$^{-10}$ in Class 0 and Class I protostars, respectively. The two values are very different between the two groups. We conducted the Kolmogorov-Smirnov (K-S) test to verify whether N(HC$_3$N)/N(N$_2$H$^+$) and x(HC$_3$N) in the two groups of Class 0 and Class I protostars are drawn from the same distribution. The K-S test gives the p-value of 0.74\% and 4.5\% for N(HC$_3$N)/N(N$_2$H$^+$) and x(HC$_3$N). The very low p-value indicates that the differences between the two groups are reliable.

We can see in the left panel of Fig. \ref{N2H_vs_HC3N} that x(N$_2$H$^+$) has a flat distribution with the N(HC$_3$N)/N(N$_2$H$^+$) ratio and displays no obvious difference between Class 0 and Class I protostars. The increasing trend of the N$_2$H$^+$ abundance can be inhibited in the center lukewarm region of the protostellar cores where the CO evaporation occurred \citep{1983A&A...122..171Y}. \citet{2018ApJ...865..135Y} also reported that the abundance of N$_2$H$^+$ begins to decrease or reaches a plateau when the dust temperature up to 27 K.
In the right panel of Fig. \ref{N2H_vs_HC3N}, it is clear that x(HC$_3$N) and N(HC$_3$N)/N(N$_2$H$^+$) increase with the evolution of the protostar. 
The increasing trend of the ratio N(HC$_3$N)/N(N$_2$H$^+$) as evolution found in high-mass protostellar objects \citep{2019ApJ...872..154T} also holds in low-mass protostellar cores with the size of less than 0.05 pc. The increased x(HC$_3$N) found here should not entirely come from sublimated CH$_4$, since the SED fitting results toward the four sources show lower dust temperatures when compared to the sublimation temperature of CH$_4$ \cite[25 K,][]{1983A&A...122..171Y}.
It is worth noting that the special Class 0 protostar shown in Fig. \ref{N2H_vs_HC3N} is L1527 that is the first WCCC source \citep{2008ApJ...672..371S}. The detection of CH$_4$ in the protoplanetary disk of protostellar in L1524 was reported by \citet{2013ApJ...776L..28G}. This serves as important evidence of the WCCC source since CH$_4$ has not been directly detected in a WCCC source thus far. Although the abundances of C$_4$H and HC$_3$N are lower than those in the WCCC source L1527, when it comes to L1524, it can be regarded as a WCCC candidate. Since I04181 has a high CCM abundance, similarly to L1527, we should also consider it a WCCC candidate.

\subsubsection{S-bearing molecules and shocked carbon-chain chemistry \label{S}}
Since CS was first detected in the interstellar medium \citep{1971ApJ...168L..53P}, S-bearing molecules have often been used to probe the physical structure of star-forming regions and sulfur chemistry \citep[e.g.,][]{1989ApJ...346..168Z,1991ApJ...368..432L,1993Ap&SS.200..183W,1999MNRAS.306..691R}. S-bearing CCMs, CCS, and CCCS are abundant in cold quiescent molecular cores and are decreasing in star-forming cores \citep[e.g.,][]{1992ApJ...392..551S,1998ApJ...506..743B,2006ApJ...642..319D,2008ApJ...678.1049S}. CCS lines were not detected or only marginally detected in low-mass star-forming regions \citep{1992ApJ...392..551S}. However, the high detection rates of CCS and CCCS were found in a CCMs survey toward embedded low-mass protostars \citep{2018ApJ...863...88L}. The abnormally high abundant CCS and CCCS are also found toward B1-a where the column densities of CCCS are larger than those of HC$_5$N \citep{2018ApJ...863...88L}. Recently, CCCS column densities were found larger than those of HC$_7$N in Ku band in three molecular outflow sources \citep{2019MNRAS.488..495W}. Shocked carbon-chain chemistry (SCCC) can explain this abnormal phenomenon \citep{2019MNRAS.488..495W}.

In our observations, CCCS (J=3--2) is detected in I04181 and L1524, while HC$_7$N is not detected in the two sources. CCCS is more abundant than HC$_7$N, which is the criterion of SCCC \citep{2019MNRAS.488..495W}.
Additionally, the CCCS abundance of I04181 (9.0$\times$10$^{-10}$) is several times as much as that of the SCCC sources \citep[2.1$\times$10$^{-10}$ for L1251A, 1.7$\times$10$^{-10}$ for IRAS 20582+7724 and 1.0$\times$10$^{-10}$ for L1221,][]{2019MNRAS.488..495W}. The CCCS abundance in L1524 (1.4$\times$10$^{-10}$) is higher than that in L1221. The abundances of sulfur atoms and ions are increasing in the shocked region, and then the abundance of CCCS increases \citep{2019MNRAS.488..495W}. Considering the detection of the outflow motions in I04181 and the association between HH 6-10 and L1524, the abundant CCCS seems to be the result of shocks and can be explained by SCCC.

The SO emission is only detected in HH211.
The increased SO abundance starts at $\sim$10$^4$ yr after the temperature up to 100 K \citep{2017MNRAS.469..435V,2018MNRAS.474.5575V}. The temperature cannot be heated up to 100 K by the central protostar on this large scale of $\sim$2$\times$10$^4$ au. The dust temperature of HH211 source is 13.6 K. 
The dynamical age less than 1000 yr for HH211 is generated by CO outflow \citep{1994ApJ...436L.189M}. 
Such a short timescale is not enough for SO to grow naturally. The SO emission detected here may be formed by S$^+$ evaporated from the grain mantles.
The molecule SO has been used as the outflow and the jet tracer for some years \citep{2005MNRAS.361..244C,2015A&A...581A..85P}.
Recently, \citet{2020MNRAS.496.2790L} investigated the G9.62+0.19 with ALMA three-millimeter observation of massive star-forming regions (ATOMS). 
\citet{2020MNRAS.496.2790L} found the correlation between SO and SiO with the principal component analysis that is supported that the two molecules trace similar shocked gas. In general, there is a complex process involved in the formation of long CCMs. Thus, it is possible that SO may be detected, while CCCS is not.

\subsection{Distributions of N-bearing inorganic molecules and CCMs \label{distributions}}
N-bearing inorganic molecules, especially NH$_3$ and N$_2$H$^+$, are often used as the dense gas tracers \citep{1999ApJS..125..161J}. NH$_3$ mapping observations toward the I04181, HH211, and L1524 regions were carried out \citep{1987A&A...173..324B,1989ApJ...341..208A,2015ApJ...805..185S}. HNC is also one of the most commonly used dense gas tracers, with a critical density of larger than 10$^4$ cm$^{-3}$.
I04181W shows three HNC emission peaks and four NH$_3$ peaks, whereas the peak positions of HNC and NH$_3$ deviate. Similar results are found in HH211NE and HH211N. The deviation among emission peaks of different molecules is very common \citep[e.g.,][]{1997ApJ...486..862P,2014A&A...569A..11S,2016ApJ...833...97K}. 

HNC and HC$_3$N (J=10--9) can be used as dense gas tracers \citep{2012MNRAS.419..238B,2020MNRAS.496.2790L}.
In our observation, HC$_3$N always emits in the HNC emission regions, as shown in I04181SE, HH211mm, HH211NW, and L1524S images. However, HC$_3$N is tentatively detected or non-detected toward I04181W, HH211N, HH211NE, and L1598 where there are also strong HNC emissions. 
For I04181SE, the size of the HNC emission structure is larger than 4$'$, while the HC$_3$N emission structure has a source size of $\sim$1$'$. With the distance adopted as 140 pc \citep[and the references therein]{2004A&A...426..503W}, the size of the HNC emission structure of I04181SE is $\sim$0.2 pc, which is consistent with the clump size of 0.1-1 pc. The substructures of HC$_3$N have sizes $\sim$0.05 pc that is similar to the typical core size \citep{2002ARA&A..40...27C}. Besides, the HC$_3$N peaks are consistent well with NH$_3$ peaks in I04181SE, HH211mm, HH211N, and HH211NW.
Thus, HC$_3$N (J=10--9) is a better tracer of dense cores than HNC (J=1--0) in such cores. 

In the L1524 region, NH$_3$ peaks at 2.5 arcmin north (L1524N) of the protostar, while HNC peaks at $\sim$1 arcmin southwest (L1524S) of the protostar. The HC$_3$N in L1524S is also stronger than L1524N. It is similar to TMC-1 where the ammonia peak (TMC-1A) and cyanopolyynes peak (TMC-1CP) are located in different places \citep{1997ApJ...486..862P}. Most of the elemental nitrogen tends to be in the form of gaseous N$_2$ and NH$_3$ as steady-state chemical models \citep{2006Natur.442..425M}. Although HNC can be formed from NH$_3$ via the pathways of NH$_3$+C$^+\rightarrow$ H$_2$NC$^+$+H and H$_2$NC$^+$+e$^-\rightarrow$ HNC+H, a high density of N$_2$ is regarded as key to the presence of abundant HNC \citep{2018MNRAS.481.4662R}.
It is thus suggested that N$_2$ and NH$_3$ are  concentrated in L1524S and L1524N, respectively.

\subsection{Co-evolution between linear hydrocarbon and cyanopolyynes in I04181SE}
As seen in the emission intensity maps of I04181 (Fig. \ref{inte-map2}), it is clear that the emissions of HC$_3$N and C$_4$H are widely spreaded in I04181SE, while the HC$_3$N emission in I04181W is concentrated in some peaks and the C$_4$H emission does not show any core-like structures.
It is consistent with the emission of HC$_7$N (J=21--20) detected by \citet{2019ApJ...871..134S}.
On the contract, the emissions of HNC and c--C$_3$H$_2$ are similar in I04181SE and I04181W.
The cyanopolyynes can be formed mainly through two pathways \citep{2016ApJ...817..147T,2018MNRAS.481.4662R}:\ in the first (pathway 1), the Nitrogen atoms in cyanopolyynes are brought in through 
reaction HCN+C$_2$H$_2$$^+$+e$^-$$\rightarrow$ HC$_3$N+H$_2$, with subsequent reactions between shorter cyanopolyynes and C$_2$H$_2$$^+$ to form longer cyanopolyynes; in the second (pathway 2), cyanopolyynes are directly synthesized from CN and unsaturated hydrocarbons such as C$_2$H$_2$. In lukewarm regions, cyanopolyynes are formed in a similar mechanism of pathway 2 but in different temperature regions, depending on the sublimation temperature of their parent species \citep{2019ApJ...881...57T}.
The chemical divergences between I04181SE and I04181W can be traced back to the
different distribution of CN, considering that c--C$_3$H$_2$ is mainly formed through C$_2$H$_2$+CH $\rightarrow$ C$_3$H$_2$+H \citep{2015ApJ...807...66Y}. 
This may explain the co-evolution between linear hydrocarbon and cyanopolyynes around I04181SE.

The C$_4$H emission is weak at the center of I04181SE, which is similar to that of HC$_7$N (J=21--20), as
detected by \citet{2019ApJ...871..134S}.
It may suggest annular distributions of hydrocarbons and cyanopolyynes in I04181SE
with non-ignorable depletion in the central dense regions \citep[e.g.,][]{2009A&A...505.1199P}.
The hydrocarbons and cyanopolyynes in I04181SE are formed from atoms and ions of C. The C and C$^+$ arise in the diffuse phase of the cloud caused by photo-dissociation and photo-ionization \citep{1992ApJ...392..551S,2008ApJ...672..371S,2019MNRAS.488..495W}. The detected C$_4$H around the protostar of I04181 should be regenerated. The regenerated C$_4$H can only be formed by sublimated CH$_4$, which results in WCCC \citep{2008ApJ...681.1385H}. However, the excitation temperatures of CCH and N$_2$H$^+$ are less than 5 K, and the dust temperature derived by Herschel data is 14.3 K. All these temperatures are too cold to sublimate CH$_4$ from the surface of the dust grain. \cite{2010ApJ...722.1633S} suggested that the WCCC occurred in the centrifugal barrier of 500-1000 au. The spatial resolution of our data is larger than 8000 au, which can not resolve the WCCC happened region. High-resolution observations are needed to confirm whether the temperature is higher than 30 K in the lukewarm region and WCCC can occur.

\section{Conclusions \label{sum}}
In this study, we carry out molecular line observations toward four low-mass outflow sources IRAS 04181+2655 (I04181), HH211, L1524, and L1598 in the 3 mm band using the PMO 13.7 m telescope and in Ku band using the SHAO 65 m TMRT. HNC (J=1--0), c--C$_3$H$_2$ (2$_{1,2}$--1$_{0,1}$), CCH (N=1--0), N$_2$H$^+$ (J=1--0), HC$_3$N (J=10--9) are detected in all four sources. The transitions J=17/2--15/2 and J=19/2--17/2 of C$_4$H are detected in I04181, HH211, and L1524, and tentatively detected in L1598. CCS (J=7--6) is detected in I04181 and marginally detected in HH211. HC$_3$N (J=2--1) is detected in I04181, HH211, and L1524. HC$_5$N (J=6--5) and CCCS (J=3--2) are detected in I04181 and L1524. SO (J=1$_2$--1$_1$) is only detected in HH211. The mapping observations of HNC (J=1--0), c--C$_3$H$_2$ (2$_{1,2}$--1$_{0,1}$), HC$_3$N (J=10--9), and C$_4$H (J=17/2--15/2) and (J=19/2--17/2) were carried out in 3 mm band. The column density of CCH was derived by RADEX code in the non-LTE analysis. For other species, the column densities were derived under the LTE assumption. The main findings are summarized as follows.

\begin{enumerate}
\item We tested the molecule pair of HC$_3$N and N$_2$H$^+$ as a chemical evolutionary indicator in low-mass star-forming cores. The two samples K-S test toward N(HC$_3$N)/N(N$_2$H$^+$) of the cores associated with Class 0 and Class I protostars, giving the p-value of 0.74\% and indicating that the differences between Class 0 and Class I protostars in N(HC$_3$N)/N(N$_2$H$^+$) are reliable. N(HC$_3$N)/N(N$_2$H$^+$) increases with evolution, which occurrs not only in high-mass protostar objects \citep{2019ApJ...872..154T} but also in low-mass star-forming cores.
\item Abundant CCMs are detected in I04181 and L1524. The CCMs abundances of I04181 are similar to the WCCC source L1527. Methane is detected in L1524 \citep{2013ApJ...776L..28G}. The two sources are regarded as WCCC candidates.
\item For the two sources of I04181 and L1524, we detect CCCS J=3--2, whereas HC$_7$N is not detected. The CCCS abundances of the two sources are larger than that of the shocked carbon-chain chemistry (SCCC) source L1221. The abundant CCCS in the two sources can be explained by SCCC. In HH211, the SO line is detected but CCCS and CCS emissions are too weak to detect. The SO may be formed by S$^+$ sublimated from the grain mantles.
\item Two filamentary structures I04181SE and HH211NW are shown in HC$_3$N (J=10--9) integrated intensity maps. The substructures of the HC$_3$N emission in I04181SE have sizes of $\sim$0.05 pc and are dense cores, indicating that HC$_3$N (J=10--9) is a good tracer of dense cores.
\item  The HC$_3$N, HC$_7$N, and C$_4$H emissions are only detected in I04181SE, while HNC and c--C$_3$H$_2$ emissions are similar between I04181SE and I04181W \citep{2019ApJ...871..134S}. The chemical divergences between the two clumps may imply different distributions for the CN radical. In I04181SE, the strong correlation between C$_4$H and HC$_3$N, HC$_7$N suggests that there is a co-evolution between linear hydrocarbon and cyanopolyynes. The annular distributions of C$_4$H and HC$_7$N indicate that the depletion cannot be ignorable in the central dense regions.
\end{enumerate}

The observations of carbon-chain molecules toward 4 low-mass star-forming cores show that I04181 and L1524 are carbon-chain rich sources and have rich sulfur-containing carbon-chain molecules. The velocity-integrated intensity maps present the circled distributions of the emissions of C$_4$H and HC$_7$N \citep[observed by][]{2019ApJ...871..134S}. It is speculated that the carbon chain molecules may dissipate in the prestellar cores and reappear in the star-forming cores. N(HC$_3$N)/N(N$_2$H$^+$) increases with evolution in low-mass protostellar cores with the size of less than 0.05 pc.

\begin{acknowledgements}
We are grateful to the staff at the Qinghai Station of PMO and Shanghai Station of SHAO for their assistance during the observations. This work was supported by the NSFC No. 11988101, 12033005, 11433008, 11373009, 11503035 and 11573036, and China Ministry of Science and Technology under State Key Development Program for Basic Research (No. 2012CB821800).
\end{acknowledgements}

\bibliographystyle{aa}

  \bibliography{39110}

\clearpage
\begin{table*}
\caption{Sources}
\label{samples}
\centering
\begin{tabular}{ccccc ccc}
\hline\hline
source & Glon & Glat & R.A.(J2000) & Decl.(J2000) & distance & type & Remark Name \\
\hline
 & deg & deg & hh mm ss & dd mm ss & kpc & & \\
\hline
IRAS 04181+2655 & 170.20399567 & -16.0245647 & 04 21 10.5 & +27 02 05.9 & 0.14 & Class I & I04181\\
HH211 & 160.48530808 & -18.00038891 & 03 43 57.1 & +32 00 50.2 & 0.32 & Class 0 & HH211\\
L1524 & 173.41692078 & -16.30599418 & 04 29 23.8 & +24 32 58.0 & 0.14 & Class I & L1524\\
L1598 & 198.72071704 & -9.16386045 & 05 52 22.4 & +08 13 34.3 & 0.9 & Class I & L1598\\
\hline
\end{tabular}
\end{table*}

\begin{table*}
\caption{Observed transitions and telescope parameters}
\label{transitions}
\centering
\resizebox{\textwidth}{60mm}{
\begin{tabular}{cccc cccc ccc}
\hline\hline
Molecule & QNs & rest freq. & $\mu$ & B & S$_{ij}$ & log$_{10}$(A$_{ij}$) & E$_{low}$ & E$_{up}$ & FWHM & $\nu_{chan}$\\
 &  & GHz  & D & MHz &  &  & K & K & $''$ & m/s\\
\hline
\multicolumn{11}{c}{Molecular line transitions observed by PMO}\\
\hline
C$_4$H & N=9-8 J=19/2-17/2 & 85.63400 & 0.9 & 4758.6 & 18.894 & -5.55477 & 16.435 & 20.546 & 53 & 210\\
... & N=9-8 J=17/2-15/2 & 85.67258 &... &... & 16.883 & -5.55765 & 16.452 & 20.563 & 53 & 210\\
c-C$_3$H$_2$ & 2$_{1,2}$-1$_{0,1}$ & 85.33889 & 3.43 & 32212 & 4.5 & -4.59289 & 2.349 & 6.445 & 53 & 210\\
CCH & N=1-0 J=3/2-1/2 F=1-1 & 87.28415 & 0.769 & 43674 & 0.246 & -6.42627 & 0.002 & 4.191 & 53 & 210\\
... & N=1-0 J=3/2-1/2 F=2-1 & 87.31692 &... &... & 2.409 & -5.65605 & 0.002 & 4.193 & 53 & 210\\
... & N=1-0 J=3/2-1/2 F=1-0 & 87.32862 &... &... & 1.2 & -5.73675 & 0.000 & 4.191 & 53 & 210\\
... & N=1-0 J=1/2-1/2 F=1-1 & 87.40200 &... &... & 1.2 & -5.73568 & 0.002 & 4.197 & 53 & 210\\
... & N=1-0 J=1/2-1/2 F=0-1 & 87.40716 &... &... & 0.482 & -5.65473 & 0.002 & 4.197 & 53 & 210\\
... & N=1-0 J=1/2-1/2 F=1-0 & 87.44651 &... &... & 0.246 & -6.42386 & 0.000 & 4.197 & 53 & 210\\ 
HNC & J=1-0 & 90.66359 & 3.05 & 45332 & 1.215 & -4.48572 & 0.000 & 4.351 & 53 & 210\\
CCS & J$_N$=7$_7$-6$_6$ & 90.68638 & 2.88 & 6477.75 & 6.953 & -4.48258 & 21.764 & 26.116 & 53 & 210\\
HC$_3$N & J=10-9 & 90.97902 & 3.73 & 4549.059  & 9.999 & -4.23748 & 19.648 & 24.014 & 53 & 210\\
N$_2$H$^+$ & J=1-0 F$_1$=1-1 & 93.17188 & 3.4 & 46586 & 3.222 & -4.40932 & 0.000 & 4.472 & 53 & 210\\
... & J=1-0 F$_1$=2-1 & 93.17370 &... &... & 5.371 & -4.40926 & 0.000 & 4.472 & 53 & 210\\
... & J=1-0 F$_1$=0-1 & 93.17613 &... &... & 1.074 & -4.40918 & 0.000 & 4.472 & 53 & 210\\
\hline
\multicolumn{11}{c}{Molecular line transitions observed by TMRT}\\
\hline
SO & J=1$_2$-1$_1$ & 13.04370 & 1.535 & 21523.556 & 1.435 & -7.53601 & 10.552 & 15.807 & 73 & 32\\
HC$_5$N & J=6-5 F=5-4 & 15.97601 & 4.33 & 1331.33 & 1.637 & -6.87816 & 1.332 & 2.684 & 60 & 27\\
... & J=6-5 F=6-5 & 15.97604 &... &... & 1.945 & -6.87581 & 1.332 & 2.684 & 60 & 27\\
... & J=6-5 F=7-6 & 15.97606 &... &... & 2.308 & -6.86356 & 1.332 & 2.684 & 60 & 27\\
CCCS & J=3-2 & 17.34225 & 3.6 & 2890.3801 & 3 & -6.44737 & 0.579 & 1.664 & 55 & 25\\
HC$_3$N & J=2-1 F=2-2 & 18.19493 & 3.73 & 4549.059 & 0.167 & -7.01213 & 0.304 & 1.310 & 52 & 24\\
... & J=2-1 F=1-0 & 18.19519 &... &... & 0.222 & -6.66538 & 0.304 & 1.310 & 52 & 24\\
... & J=2-1 F=2-1 & 18.19627 &... &... & 0.5 & -6.53500 & 0.304 & 1.310 & 52 & 24\\
... & J=2-1 F=3-2 & 18.19627 &... &... & 0.933 & -6.41003 & 0.304 & 1.310 & 52 & 24\\
... & J=2-1 F=1-2 & 18.1970763 &... &... & 0.011 & -7.96613 & 0.304 & 1.310 & 52 & 24\\
... & J=2-1 F=1-1 & 18.19836 &... &... & 0.167 & -6.79020 & 0.304 & 1.310 & 52 & 24\\
\hline
\end{tabular}}
\tablefoot{The transitions of CH$_3$CCH J=5--4 (85.45727 GHz) and HC$_7$N J=14--13 (15.791987 GHz), J=15--14 (16.9199791 GHz), and J=16--15 (18.0479697 GHz) are observed but not detected, and thus unlisted. All the transitions and their properties are obtained from the Cologne Database for Molecular Spectroscopy\footnote{http://www.astro.uni-koeln.de/cdms/}.}
\end{table*}

\clearpage
\begin{table*}
  \centering
  \caption{Line parameters of PMO}
  \label{line-para-PMO}
  \centering
  \resizebox{\textwidth}{70mm}{
  \begin{tabular}{ccccc ccccc}
\hline
\hline
&  & \multicolumn{4}{c}{I04181} &  \multicolumn{4}{c}{HH211}\\
  \cmidrule(r){3-6} \cmidrule(r){7-10}
  molecules & transitions & T$_a$ & V$_{lsr}$ & $\Delta$V & T$_a$*$\Delta$V & T$_a$ & V$_{lsr}$ & $\Delta$V & T$_a$*$\Delta$V \\
   &  & K & km s$^{-1}$ & km s$^{-1}$ & K km s$^{-1}$ & K & km s$^{-1}$ & km s$^{-1}$ & K km s$^{-1}$ \\
\hline
C$_4$H & J=17/2-15/2 & 0.15(0.08) & 6.73(0.07) & 0.56(0.15) & 0.09(0.03) & 0.18(0.06) & 9.11(0.03) & 0.41(0.08) & 0.08(0.02) \\
C$_4$H   & J=19/2-17/2 & 0.13(0.08) & 6.81(0.06) & 0.44(0.14) & 0.06(0.03) & 0.10(0.04) & 9.09(0.10) & 1.42(0.27) & 0.15(0.03) \\
HNC      & J=1-0   & 0.71(0.04) & 7.06(0.01) & 0.95(0.03) & 0.72(0.06) & 1.23(0.02) & 8.81(0.01) & 1.23(0.01) & 1.61(0.03) \\
HC$_3$N  & J=10-9  & 0.47(0.06) & 6.76(0.01) & 0.49(0.03) & 0.24(0.04) & 0.35(0.05) & 9.04(0.01) & 0.47(0.03) & 0.18(0.03) \\
CCS      & J$_N$=7$_7$-6$_6$   & 0.10(0.04) & 6.81(0.06) & 0.56(0.13) & 0.06(0.01) & 0.06(0.04) & 9.09(0.07) & 0.36(0.11) & 0.02(0.01) \\
c-C$_3$H$_2$ & 2$_{1,2}$-1$_{0,1}$ & 0.66(0.06) & 6.86(0.01) & 0.51(0.03) & 0.36(0.02) & 0.49(0.06) & 9.10(0.01) & 0.58(0.04) & 0.30(0.02) \\
CCH & N=1-0 J=3/2-1/2 F=1-1   & 0.19(0.06) & ...        & 0.61(0.09) & 0.12(0.05) & 0.17(0.08) & ...        & 0.97(0.22) & 0.17(0.12) \\
    & N=1-0 J=3/2-1/2 F=2-1   & 0.72(0.08) & 6.93(0.01) & 0.61(0.03) & 0.47(0.07) & 0.79(0.14) & 9.02(0.02) & 0.74(0.07) & 0.62(0.16) \\
    & N=1-0 J=3/2-1/2 F=1-0   & 0.51(0.07) & ...        & 0.63(0.04) & 0.34(0.06) & 0.58(0.21) & ...        & 0.54(0.11) & 0.33(0.18) \\
    & N=1-0 J=1/2-1/2 F=1-1   & 0.57(0.06) & ...        & 0.58(0.05) & 0.35(0.06) & 0.58(0.18) & ...        & 0.72(0.12) & 0.45(0.20) \\
    & N=1-0 J=1/2-1/2 F=0-1   & 0.36(0.12) & ...        & 0.63(0.11) & 0.24(0.12) & 0.27(0.13) & ...        & 1.01(0.27) & 0.29(0.20) \\
    & N=1-0 J=1/2-1/2 F=1-0   & 0.24(0.10) & ...        & 0.43(0.08) & 0.11(0.06) & 0.21(0.25) & ...        & 0.68(0.49) & 0.15(0.27) \\
N$_2$H$^+$ & J=1-0 F$_1$=1-1  & ... & 6.78(0.09) & ...  & 0.19(0.04) & ...  & 8.76(0.07) & ...   & 0.63(0.04) \\
& J=1-0 F$_1$=2-1  & ...   & ...  & ...  & 0.29(0.06) & ...   & ...  & ...   & 0.85(0.08)  \\
& J=1-0 F$_1$=0-1 & 0.12(0.03) & ... & 0.40(0.20) & 0.05(0.02) & 0.30(0.03) & ...   & 0.67(0.23) & 0.22(0.06)  \\
\hline
  & & \multicolumn{4}{c}{L1524} & \multicolumn{4}{c}{L1598}\\
\cmidrule(r){3-6} \cmidrule(r){7-10}
C$_4$H & J=17/2-15/2  & 0.33(0.10) & 6.20(0.01) & 0.21(0.11) & 0.07(0.02) & 0.10(0.06) & 11.27(0.05) & 0.38(0.13) & 0.04(0.02)\\
C$_4$H & J=19/2-17/2  & 0.21(0.08) & 6.26(0.04) & 0.47(0.10) & 0.10(0.03) & 0.10(0.06) & 11.36(0.05) & 0.42(0.13) & 0.05(0.02)\\
HNC    & J=1-0    & 0.56(0.03) & 6.38(0.02) & 1.17(0.03) & 0.70(0.04) & 0.64(0.02) & 11.17(0.01) & 1.35(0.02) & 0.92(0.03)\\
HC$_3$N  & J=10-9  & 0.33(0.05) & 6.21(0.02) & 0.50(0.04) & 0.17(0.03) & 0.12(0.07) & 11.27(0.03) & 0.34(0.12) & 0.04(0.03)\\
CCS  & J$_N$=7$_7$-6$_6$   & ...  & ... & ... & ...  & ...  & ... & ...  & ...  \\
c-C$_3$H$_2$ & 2$_{1,2}$-1$_{0,1}$ & 0.87(0.09) & 6.29(0.01) & 0.50(0.03) & 0.47(0.02) & 0.26(0.04) & 11.23(0.03) & 0.81(0.07) & 0.23(0.02)\\
CCH & N=1-0 J=3/2-1/2 F=1-1   & 0.38(0.13) & ...        & 0.36(0.07) & 0.15(0.07) & ...        & ...         & ...        & ...       \\
    & N=1-0 J=3/2-1/2 F=2-1   & 1.07(0.10) & 6.33(0.01) & 0.58(0.03) & 0.66(0.09) & 0.32(0.24) & 11.33(0.05) & 0.95(0.13) & 0.32(0.27)\\
    & N=1-0 J=3/2-1/2 F=1-0   & 0.71(0.21) & ...        & 0.54(0.09) & 0.41(0.18) & 0.19(0.08) & ...         & 1.01(0.20) & 0.20(0.12)\\
    & N=1-0 J=1/2-1/2 F=1-1   & 0.84(0.11) & ...        & 0.56(0.04) & 0.50(0.10) & 0.18(0.07) & ...         & 1.13(0.24) & 0.21(0.12)\\
    & N=1-0 J=1/2-1/2 F=0-1   & 0.51(0.18) & ...        & 0.47(0.08) & 0.26(0.12) & ...        & ...         & ...        & ...       \\
    & N=1-0 J=1/2-1/2 F=1-0   & 0.41(0.12) & ...        & 0.33(0.05) & 0.14(0.05) & ...        & ...         & ...        & ...       \\
N$_2$H$^+$ & J=1-0 F$_1$=1-1   & ... & 6.24(0.04) & ...   & 0.29(0.05) & ...  & 10.97(0.09) & ...  & 0.44(0.08)\\
 & J=1-0 F$_1$=2-1   & ... & ...   & ...    & 0.37(0.05) & ...   & ...   & ...  & 0.52(0.10) \\
& J=1-0 F$_1$=0-1   & 0.24(0.03) & ...  & 0.26(0.10) & 0.07(0.02) & 0.22(0.03) & ...   & 0.74(0.22) & 0.17(0.04) \\
\hline
\end{tabular}}
\tablefoot{The numbers in parentheses stand for the standard deviation of Gaussian fitting except the errors of integrated intensities of two groups F$_1$=1--1 and F$_1$=2--1 of N$_2$H$^+$ (J=1--0) which is estimated as $T_{rms}\times\sqrt{N_{channels}\times\delta V_{res}}$.}
\end{table*}

\begin{table*}[htp!]
  \caption{Line parameters of TMRT}
  \label{line-para-TMRT}
\begin{tabular}{ccc ccc}
  \hline\hline
molecules & transitions & T$_a$ & V$_{lsr}$ & $\Delta$V & T$_a$*$\Delta$V\\
    &   & K & km s$^{-1}$ & km s$^{-1}$ & K km s$^{-1}$\\
\hline
\multicolumn{6}{c}{I04181}\\
\hline
HC$_3$N &  J=2-1 F=1-1  & 0.305(0.096) & ... & 0.654(0.148) & 0.212(0.040)\\
& F=3-2  & 0.741(0.102) & 6.817(0.021) & 0.499(0.049) & 0.394(0.033)\\
& F=2-1  & 0.434(0.112) & ... & 0.465(0.087) & 0.215(0.033)\\
& F=1-0  & 0.216(0.120) & ... & 0.453(0.186) & 0.104(0.034)\\
HC$_5$N &  J=6-5 F=6--5? & 0.153(0.042) & 6.528(0.045) & 0.529(0.100) & 0.086(0.015)\\
& F=5-4?  & 0.069(0.034) & ... & 0.758(0.258) & 0.056(0.017)\\
CCCS &  J=3-2        & 0.152(0.040) & 6.857(0.032) & 0.416(0.079) & 0.067(0.011)\\
\hline
\multicolumn{6}{c}{HH211}\\
\hline
HC$_3$N &  J=2-1 F=3-2  & 0.481(0.109) & 9.231(0.029) & 0.444(0.074) & 0.227(0.030)\\
SO &  J=1$_2$-1$_1$ & 0.212(0.026) & 8.714(0.022) & 0.544(0.045) & 0.123(0.010)\\
\hline
\multicolumn{6}{c}{L1524}\\
\hline
HC$_3$N &  J=2-1 F=1-1  & 0.447(0.144) & ... & 0.215(0.050) & 0.102(0.020)\\
& F=3-2  & 1.251(0.172) & 6.313(0.016) & 0.370(0.036) & 0.493(0.042)\\
& F=2-1  & 0.595(0.098) & ...  & 0.394(0.045) & 0.250(0.026)\\
& F=1-0  & 0.364(0.104) & ... & 0.443(0.093) & 0.172(0.029)\\
& F=2-2  & 0.234(0.118) & ...  & 0.252(0.088) & 0.063(0.020)\\
HC$_5$N &  J=6-5 F=?      & 0.154(0.036) & 6.136(0.037) & 0.518(0.089) & 0.085(0.012)\\
CCCS &  J=3-2        & 0.156(0.054) & 6.329(0.023) & 0.249(0.065) & 0.041(0.008)\\
\hline
\end{tabular}
\tablefoot{The numbers in parentheses stand for the standard deviation of Gaussian fitting.}
\end{table*}

\begin{table*}[htp!]
  \caption{Hyperfine structure fitting parameters}
  \label{hfs}
\begin{tabular}{cccc cc}
\hline\hline
source & T$_a$*$\tau$ & V$_{lsr}$ & $\Delta$V & $\tau$ & T$_{ex}$ \\
 & K & km s$^{-1}$ & km s$^{-1}$ &   & K \\
\hline
\multicolumn{6}{c}{HC$_3$N (J=2--1)}\\
\hline
L1524 & 1.46(0.186) & 6.31(0.008) & 0.348(0.020) & 0.432(0.338) & 8.7(3.9)\\
\hline
\multicolumn{6}{c}{CCH (N=1--0)}\\
\hline
I04181 & 1.36(0.09) & 6.91(0.01) & 0.55(0.02) & 1.46(0.23) & 4.8(0.3)\\
HH211 & 1.09(0.10) & 9.01(0.02) & 0.73(0.05) & 0.67(0.24) & 6.4(1.2)\\
L1524 & 1.82(0.12) & 6.33(0.01) & 0.50(0.02) & 1.19(0.23) & 6.0(0.4)\\
L1598 &  0.33(0.02) & 11.32(0.04) & 1.04(0.09) & 2.32(1.22) & ... \\
\hline
\multicolumn{6}{c}{N$_2$H$^+$ (J=1--0)}\\
\hline
I04181 & 0.28(0.08)  & 6.85(0.02) &  0.51(0.06) &  0.46(0.69) & 3.4(1.6) \\
HH211  & 1.55(0.29)  & 8.99(0.02) &  0.44(0.03) &  2.68(0.84) & 6.1(2.1) \\
L1524  & 0.74(0.19)  & 6.25(0.02) &  0.34(0.04) &  1.72(0.95) & 4.4(2.1) \\
L1598  & 1.18(0.43)  & 11.30(0.03) &  0.50(0.07) &  4.12(1.98) & 5.4(2.6) \\
\hline
\end{tabular}
\tablefoot{The numbers in parentheses stand for the standard deviation of the HFS fitting.}
\end{table*}

\begin{table*}
  \centering
  \caption{Column densities of molecules\tablefootmark{\,{\footnotesize a}}}
  \label{density}
  \resizebox{\textwidth}{12mm}{
\begin{tabular}{ccc ccc ccc ccc}
  \hline\hline
  Source & T$_{\mathrm{dust}}$ & CCH & N$_2$H$^+$ & C$_4$H & HC$_3$N J=10--9 & CCS & 
         c--C$_3$H$_2$ & HC$_3$N J=2-1 & HC$_5$N & CCCS & SO \\
    & K &  E+13 (cm$^{-2}$) &  E+12 (cm$^{-2}$) &    E+13 (cm$^{-2}$) &  E+12 (cm$^{-2}$)  &   E+12 (cm$^{-2}$) &  E+12 (cm$^{-2}$) &   E+13 (cm$^{-2}$) &  E+13 (cm$^{-2}$) &   E+12 (cm$^{-2}$) &  E+13 (cm$^{-2}$)\\
  \hline
  I04181 & 14.32 & 9.7(25) & 3.2(5)   & 1.4(5) & 2.8(3) & 4.1(7) & 1.4(1)  & 5.0(4) & 1.6(3) & 4.5(7) & ... \\
  HH211  & 13.61 & 9.8(13) & 15.4(21) & 2.1(4) & 2.2(4) & <1.4   & 1.2(1)  & 2.8(4) & ... & ... & 5.0(4) \\
  L1524  & 15.76 & 10.1(25) & 5.1(9)   & 1.4(4) & 1.8(3) & ...    & 2.0(1)  & 6.7(6) & 1.7(2) & 2.9(6) & ... \\
  L1598  & 15.16 & 3.2(23) & 17(21)   & 0.8(4) & <0.4   & ...    & 0.9(1)  & ...    & ... & ... & ...\\
  \hline
\end{tabular}}
\tablefoot{The numbers in parentheses represent the errors in units of the last significant digits.}
\tablefoottext{a}{The column densities of CCH are derived by RADEX code. The column densities of N$_2$H$^+$ are derived by T$_{ex}$ obtained by HFS fitting.}
\end{table*}

\begin{table*}
  \centering
  \caption{Column density ratios}
  \label{ratios}
  \resizebox{\textwidth}{11mm}{
\begin{tabular}{ccc cccc}
  \hline\hline
  source  & N(CCH)/N(N$_2$H$^+$) & N(HC$_3$N J=10--9)/N(N$_2$H$^+$)  & N(C$_4$H)/N(N$_2$H$^+$) & N(CCH)/N(C$_4$H) & N(HC$_3$N J=2--1)/N(HC$_5$N) &  N(HC$_5$N)/N(CCCS)\\
  \hline
  I04181 &   30(16)  & 0.87(23)   & 4.4(22)  &  7(5)    & 3.1(8)     & 3.6(12) \\
  HH211 &    7(2)  & 0.14(4)    & 1.4(4) &  4(2)  & ... & ...\\
  L1524 &    19(9)  & 0.35(12)   & 2.7(13) &  7(4)   & 3.9(8)     & 5.9(19) \\
  L1598 &    2(2)  & 0.03(1)   & 0.6(3) &  4(6)   & ... & ...\\
  \hline
\end{tabular}}
\tablefoot{The numbers in parentheses represent the errors in units of the last significant digits.}
\end{table*}
  
\begin{table*}
  \centering
  \caption{Abundances}
  \label{abundance}
\begin{tabular}{ccc ccc c ccc c}
  \hline\hline
  Source & C$_4$H & HC$_3$N J=10--9 & CCS & c--C$_3$H$_2$ & CCH & 
  N$_2$H$^+$ & HC$_3$N J=2--1 & HC$_5$N & CCCS & SO \\
    &   E-9  & E-10 &   E-10  &  E-10  &   E-9  &  E-10  &   E-9 &  E-9  &   E-10 &  E-9\\
  \hline
  I04181 & 2.8(11) & 5.6(7) & 8.2(15) & 2.8(2) & 19.4(69)       & 6.4(10)   & 10.0(9) & 3.2(7) & 9.0(15) & ... \\
  HH211 &  0.4(1) & 0.4(1) & <0.7 & 0.2(0.2) & 1.7(4)        & 2.6(3)  & 0.5(1) & ... & ... & 0.8(1) \\
  L1524 &  0.7(2) & 0.9(2) & ... & 1.0(1) & 5.0(17)        & 2.6(3)  & 3.3(4) & 0.8(1) & 1.4(3) & ... \\
  L1598 &  0.7(4) & <0.4 & ... & 0.8(1) & 3.0(24)       & 15.5(19)  & ... & ... & ... & ...\\
  \hline
\end{tabular}
\tablefoot{The numbers in parentheses represent the errors in units of the last significant digits.}
\end{table*}

\begin{table*}
  \caption{Summary of statistical analyses}
  \label{stt}
  \begin{tabular}{ccc}
    \hline\hline
  & N(HC$_3$N J=10--9)/N(N$_2$H$^+$) & x(HC$_3$N)\\
  \hline
  K-S test p-value & 0.74\%  & 4.50\%  \\
  mean(Class 0) & 0.43  &3.42$\times$10$^{-10}$\\
  mean(Class I) & 2.55  &20.18$\times$10$^{-10}$\\
  \hline
  \end{tabular}
\end{table*}

\clearpage

\clearpage
\begin{appendix}

\section{Greybody SED fitting}
All the observed sources have \emph{Herschel} data at 70 $\mu$m, 160 $\mu$m, 250 $\mu$m, 350 $\mu$m, and 500 $\mu$m -- except L1598, which has 100 $\mu$m replacing the 70 $\mu$m data in the \emph{Herschel Science Archive}\footnote{http://www.cosmos.esa.int/hsa/whsa/}, from which we extracted the Level 2 images.
Thus, we selected the wavelengths of 160 $\mu$m, 250 $\mu$m, 350 $\mu$m, and 500 $\mu$m to estimate dust temperatures and column densities. These data have half-power beam widths (HPBWs) of 13.5, 18.1, 24.9, and 36.4 arcsec at 160 $\mu$m, 250 $\mu$m, 350 $\mu$m, and 500 $\mu$m, respectively \citep{2011MNRAS.416.2932T}.

Background and foreground emission were removed to follow the \emph{CUPID-findback} algorithm\footnote{http://starlink.eao.hawaii.edu/starlink/WelcomePage} \citep{2007ASPC..376..425B}. The background is determined using a filtering process as the following three steps. In the first step, a filtered form of the input data is produced by replacing every pixel with the minimum of the input values within a square box centered on the pixel. The square box defined in this work is 5$'\times$5$'$. In the second step, these filtered data are filtered again, using a filter that replaces every pixel value by the maximum value in the box centered on the pixel. In the third step, the filtered data produced by in step two are filtered again by replacing each value by the mean value in the box centered on the value being replaced.

\begin{figure}[h]
  \renewcommand{\thefigure}{A1}
  \centering
  \includegraphics[width=0.9\linewidth]{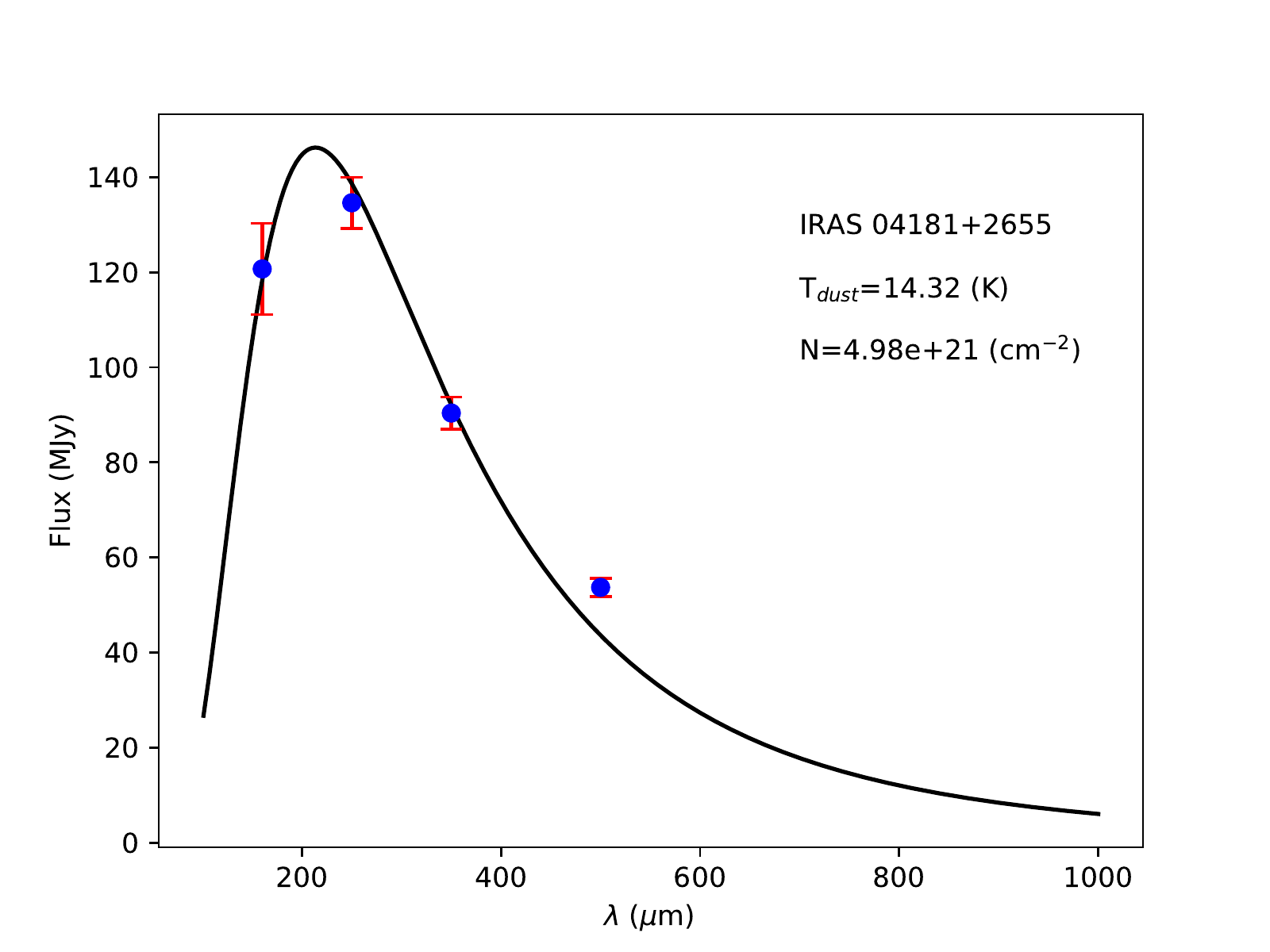}\\
  \includegraphics[width=0.9\linewidth]{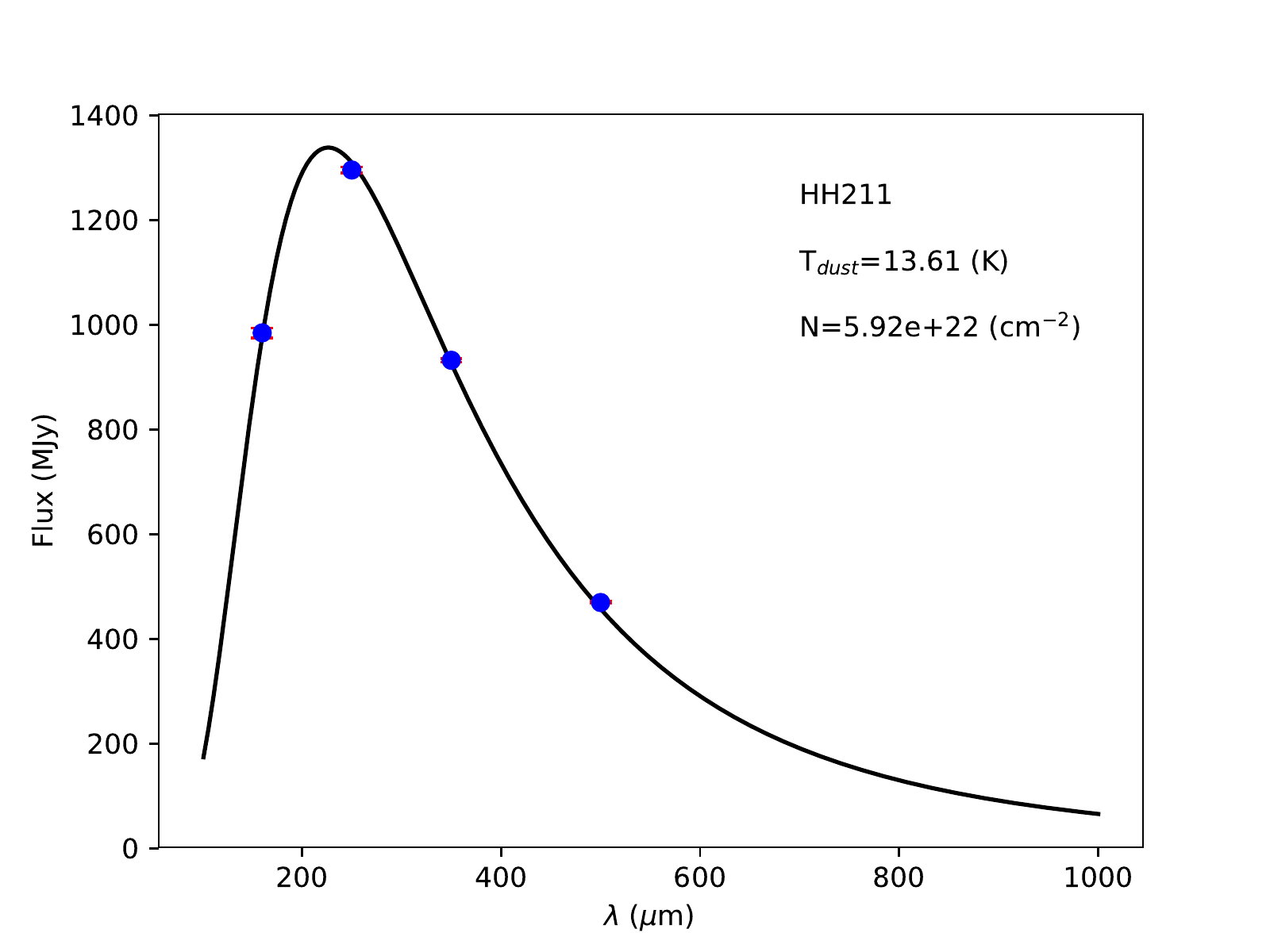}\\
  \includegraphics[width=0.9\linewidth]{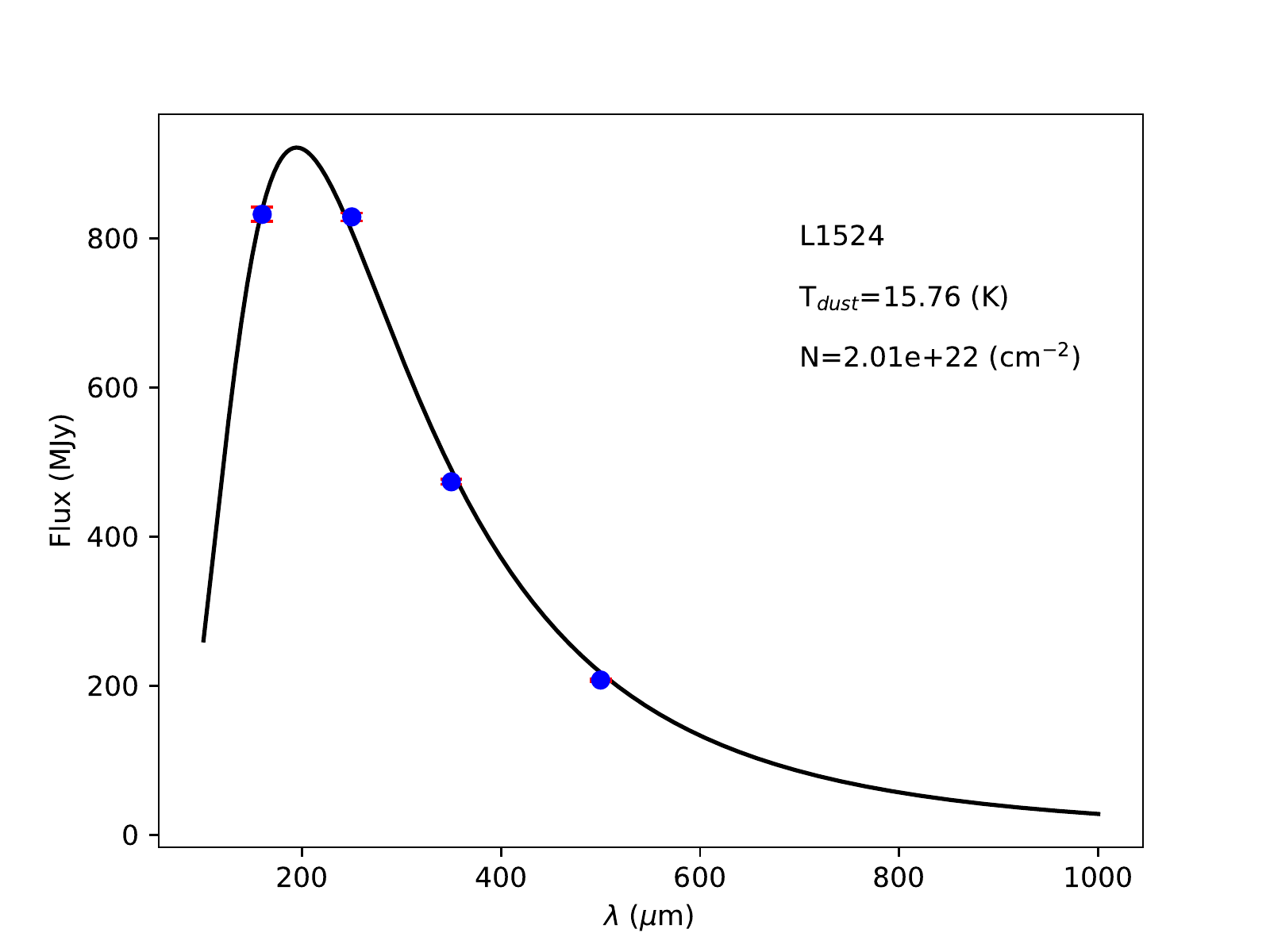}\\
  \includegraphics[width=0.9\linewidth]{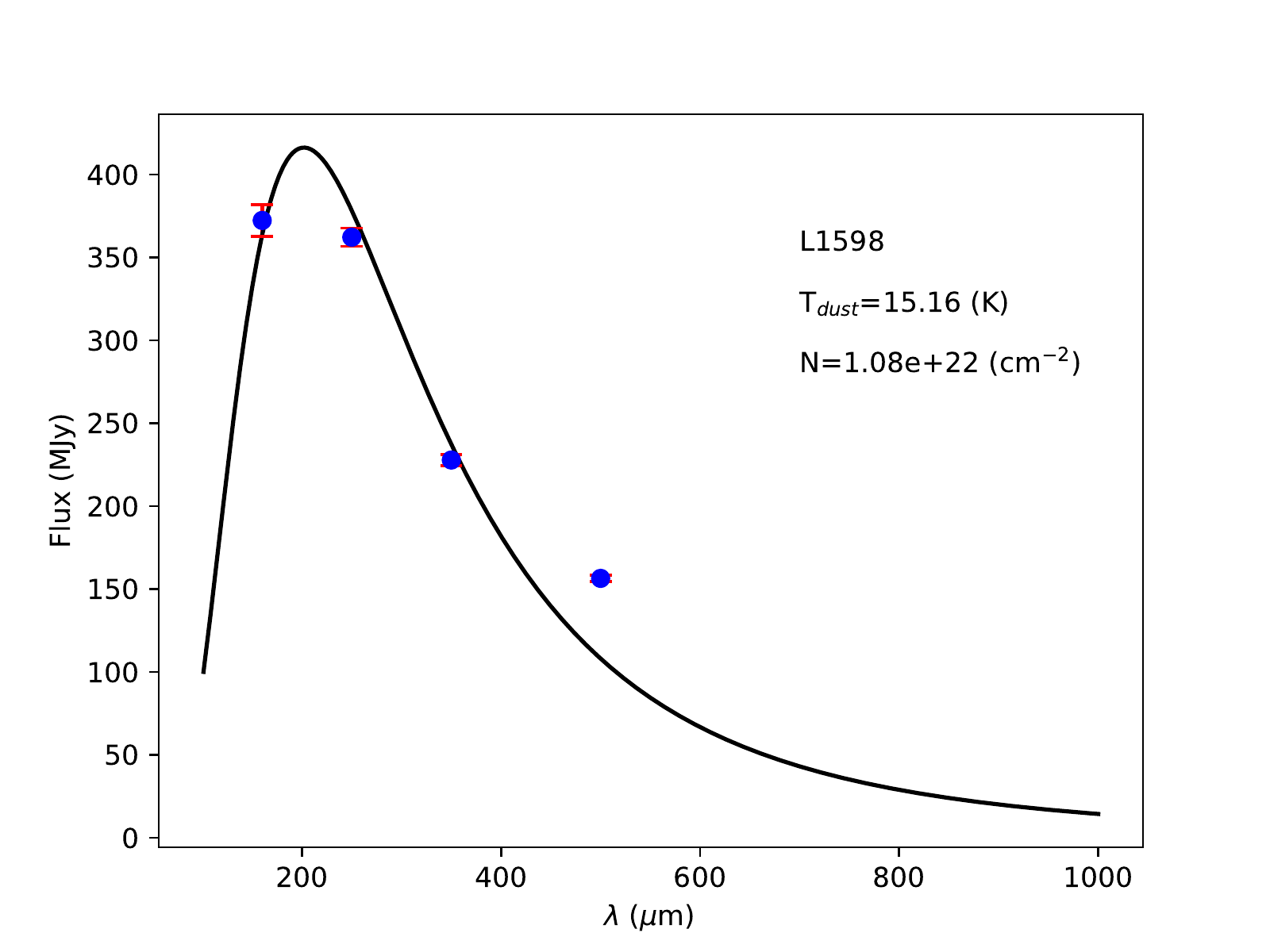}
  \caption{SEDs of the observed sources. The blue dots are the data points used for the SED fitting. The error bar of each point is plotted. \label{SED}}
\end{figure}

Then we convolved each waveband image with a Gaussian Kernel with HPBW to 36.4 arcsec the 500 $\mu$m beam size.
Pixel-by-pixel greybody SED fittings were applied:\ 
\begin{eqnarray}
  I_{\nu} = B_{\nu}(T_{dust})(1-e^{-\tau_{\nu}}),
\end{eqnarray} 
where the $B_{\nu}(T_{dust})$ is the Planck function, the dust optical depth $\tau_{\nu}$ can be expressed as: 
\begin{eqnarray}
  \tau_{\nu} = \mu_{H_2}m_H \kappa_{\nu} N_{H_2}/R_{gd}.
\end{eqnarray} 
Here, $\mu_{H_2}$ is the mean weight of molevule which equals to 2.33 \citep{2008A&A...487..993K}, m$_H$ is the mass of a hydrogen atom, N$_{H_2}$ is the column density, and R$_{gd}$ is the gas to dust raito with value assumed as 100. The dust opacity $\kappa_{\nu}$ can be expressed as 
\begin{eqnarray}
  \kappa_{\nu} = \kappa_0\left(\frac{\nu}{\nu_0}\right)^{\beta},
\end{eqnarray} 
where $\kappa_0$ is given as 1.14 cm$^2$g$^{-1}$ at $\nu_0$ of 271.1 GHz by \citet{1994A&A...291..943O}. The dust emissivity index $\beta$ is fixed as 1.75. The free parameters are the dust temperature and column density. The SEDs of single points toward the four sources are plotted as shown in Fig. \ref{SED}.

\end{appendix}

\end{CJK}
\end{document}